\documentclass[fleqn,usenatbib]{mnras}
\usepackage[T1]{fontenc}
\DeclareRobustCommand{\VAN}[3]{#2}
\let\VANthebibliography\thebibliography
\def\thebibliography{\DeclareRobustCommand{\VAN}[3]{##3}\VANthebibliography}
\usepackage{graphicx}	
\usepackage{subcaption}
\usepackage{amsmath}	
\usepackage{amssymb}	
\usepackage[dvipsnames]{xcolor}
\usepackage{fontawesome}

\usepackage{newtxtext,newtxmath}

\title[\textsc{CosmoPower}]{\textsc{CosmoPower}: emulating cosmological power spectra for accelerated Bayesian inference from next-generation surveys}

\author[A. Spurio Mancini et al.]{
Alessio Spurio Mancini$^{1, 2}$\thanks{\href{mailto:a.spuriomancini@ucl.ac.uk}{a.spuriomancini@ucl.ac.uk}},
Davide Piras$^{2}$,
Justin Alsing$^{3}$,
Benjamin Joachimi$^{2}$,
Michael P. Hobson$^{4}$
\\
$^{1}$Mullard Space Science Laboratory, University College London,
Holmbury St. Mary, Dorking, Surrey, RH5 6NT, UK \\
$^{2}$Department of Physics and Astronomy, University College London, Gower Street, London, WC1E 6BT, UK \\
$^{3}$Oskar Klein Centre for Cosmoparticle Physics, Department of Physics, Stockholm University, Stockholm SE-106 91, Sweden \\
$^{4}$Astrophysics Group, Cavendish Laboratory, J. J. Thomson Avenue, Cambridge, CB3 0HE, UK
}

\date{Accepted XXX. Received YYY; in original form ZZZ}

\pubyear{2021}

\begin{document}
\label{firstpage}
\pagerange{\pageref{firstpage}--\pageref{lastpage}}
\maketitle

\begin{abstract}
We present \textsc{CosmoPower}, a suite of neural cosmological power spectrum emulators providing orders-of-magnitude acceleration for parameter estimation from two-point statistics analyses of Large-Scale Structure (LSS) and Cosmic Microwave Background (CMB) surveys. The emulators replace the computation of matter and CMB power spectra from Boltzmann codes; thus, they do not need to be re-trained for different choices of astrophysical nuisance parameters or redshift distributions. The matter power spectrum emulation error is less than $0.4\%$ in the wavenumber range $k \in [10^{-5}, 10] \, \mathrm{Mpc}^{-1}$, for redshift $z \in [0, 5]$. \textsc{CosmoPower} emulates CMB temperature, polarisation and lensing potential power spectra in the $5\sigma$ region of parameter space around the \textit{Planck} best fit values with an error $\lesssim 10\%$ of the expected shot noise for the forthcoming Simons Observatory. \textsc{CosmoPower} is showcased on a joint cosmic shear and galaxy clustering analysis from the Kilo-Degree Survey, as well as on a Stage IV \textit{Euclid}-like simulated cosmic shear analysis. For the CMB case, \textsc{CosmoPower} is tested on a \textit{Planck} 2018 CMB temperature and polarisation analysis. The emulators always recover the fiducial cosmological constraints with differences in the posteriors smaller than sampling noise, while providing a speed-up factor up to $O(10^4)$ to the complete inference pipeline. This acceleration allows posterior distributions to be recovered in just a few seconds, as we demonstrate in the \textit{Planck} likelihood case. \textsc{CosmoPower} is written entirely in \textsc{Python}, can be interfaced with all commonly used cosmological samplers and is publicly available \href{https://github.com/alessiospuriomancini/cosmopower}{\faicon{github}}.
\end{abstract}

\begin{keywords}
large-scale structure of Universe -- cosmic background radiation -- methods: statistical -- methods: data analysis
\end{keywords}


\section{Introduction}\label{sec:introduction}
Analysis of the two-point statistics of cosmological fields is one of the cornerstones of modern observational cosmology. For parameter inference pipelines involving two-point statistics (i.e. power spectra, or their derived real-space counterparts, correlation functions), the computational bottleneck is running Boltzmann solvers like \textsc{Camb} \citep{Lewis99} or \textsc{Class} \citep{Lesgourgues11, Blas11} to compute theoretical power spectra for a given cosmology. However, cosmological power spectra are generally smooth functions of their input cosmological parameters, and hence lend themselves well to \emph{emulation}: finding compact, accurate, and fast-to-evaluate surrogate functions that map cosmological parameters to the corresponding predicted power spectra. 

Emulation offers the promise of reducing the computational overhead of evaluating cosmological power spectra by many orders of magnitude, with negligible loss of accuracy in the final parameter inference. This surrogate modelling approach has recently seen numerous applications to the Bayesian solution of the \textit{inverse problem} in different scientific fields, ranging from geophysical seismic waves \citep{Das18, SpurioMancini20, Piras21} to stellar and galaxy spectra \citep{Czekala15, Alsing20}, from chemical mechanisms \citep{deMijolla19, Kasim20} to applied engineering \citep{Thiagarajan20, Buffington20}.

Emulation of cosmological power spectra is not a new idea either. Early examples of emulators include \textsc{CMBwarp} \citep{Jimenez04} and \textsc{PICO} \citep{Fendt07}, both performing polynomial regression of power spectra (represented in some compact basis). The matter power spectrum emulators built from the Coyote Universe simulations \citep{Heitmann09, Heitmann10, Heitmann13, Lawrence10b, Lawrence17} are based on Gaussian Process regression \citep{Rasmussen05}, and were extended by \citet{Ramachandra20} to $f(R)$ cosmologies. Recently, the Euclid Emulator was proposed as a surrogate model for the nonlinear matter power spectrum \citep{Knabenhans19, euclidcollaboration20}, \citet{Mootoovaloo20} developed a Gaussian Process emulator of cosmic shear band powers, while \citet{Mootoovaloo21} and \citet{Ho21} used Gaussian Processes to emulate the matter power spectrum. \citet{Bird19} and \citet{Rogers19} developed Gaussian Process emulators for the Lyman-$\alpha$ forest flux power spectrum.

\textsc{CosmoNet} \citep{Auld07, Auld08} is the first application of neural networks to cosmological power spectra emulation, followed by \citet{Agarwal12, Agarwal14}. More recently, \citet{ManriqueYus19} developed neural network interpolators for angular power spectra of LSS observables, while \citet{Albers19} used neural networks to accelerate parts of power spectra computations within the Boltzmann code \textsc{Class}; moreover, the \textsc{Bacco} project \citep{Angulo20} recently included a neural network interpolator for the linear matter power spectrum \citep{Arico21}. \citet{Kern17}, \citet{Schmit17} and \citet{Bevins21} developed neural network emulators for the 21-cm global signal or power spectrum. 

In this paper, we introduce a suite of neural cosmological power spectrum emulators covering both CMB (temperature, polarisation and lensing) and (linear and nonlinear) matter power spectra. These emulators provide orders-of-magnitude speed-up over direct Boltzmann solvers, whilst being comfortably accurate for upcoming surveys such as the Simons Observatory \citep{Ade19}\footnote{\url{https://simonsobservatory.org/}}, \textit{Euclid} \citep{Laureijs11}\footnote{\url{https://www.euclid-ec.org/}}, the Vera Rubin Observatory \citep{Ivezic19}\footnote{\url{https://www.lsst.org/}} and the Nancy Grace Roman Space Telescope \citep{Spergel15}\footnote{\url{https://roman.gsfc.nasa.gov/}}. For LSS observables, we demonstrate the accuracy and acceleration provided by our emulators on data from the Kilo-Degree Survey (KiDS) as well as on a simulated cosmic shear analysis of a \textit{Euclid}-like survey. For the CMB, we validate \textsc{CosmoPower} on a \textit{Planck} 2018 CMB temperature and polarisation analysis.

Our emulators are trained to provide accurate emulation of cosmological power spectra on a very wide range of cosmological parameters, and they easily allow for different configurations of input and derived cosmological parameters. In addition, they are fully differentiable, which makes them amenable to gradient-based inference, and can be run on Graphics Processing Units to gain further acceleration. 

The structure of this paper is as follows. In Sect.~\ref{sec:methods} we introduce the neural network emulation framework tailored to power spectrum emulation. Application to the matter power spectrum emulation is presented in Sect.~\ref{sec:LSS}, including validation on the KiDS and \textit{Euclid}-like analyses. Application to the CMB case is presented in Sect.~\ref{sec:CMB}, including validation on the \textit{Planck} analysis. We summarise the properties of \textsc{CosmoPower} and conclude in Sect.~\ref{sec:conclusions}.

\section{Emulating cosmological power spectra}\label{sec:methods}
We consider two methods for power spectra emulation.
\begin{itemize}
    \item The first one is a direct mapping between cosmological parameters and power spectra by means of a neural network (NN), schematically represented in the left-hand panel of Fig.~\ref{fig:emulation} (see e.g. \citealt{Bishop06} for an introduction to NNs; here, we follow a similar notation to \citealt{Alsing20}). A NN is a set of stacked layers, each composed of multiple nodes; we label each of the $n$ nodes with index $i$. To each of them a weight $W_i$ and bias parameter $b_i$ are associated, whose linear combination is passed through a nonlinear activation function (see Appendix \ref{app:nn} for details on the activation function employed in this work); we denote the ensemble of weights and biases as $\boldsymbol{w}$. The nonlinear function parameterised by $\boldsymbol{w}$ maps the input cosmological parameters $\boldsymbol{\theta}$ to the (log-)power spectra $\boldsymbol{P}_\lambda=\boldsymbol{P}_\lambda(\boldsymbol{\theta}; \boldsymbol{w})$, where $\lambda$ is either the wavenumber $k$ for the matter power spectrum, or the multipole $\ell$ in the CMB case. The spectra used for training the deep learning emulators are preprocessed by calculating their logarithm, followed by standardisation of the logarithmic features, i.e. dividing each logarithmic feature by its standard deviation after subtracting its mean. Taking the logarithm of the spectra reduces the dynamic range in the training data, and ensures that minimising the mean-square-error loss optimises for fractional (rather than absolute) accuracy on the emulated power spectrum. Standardisation ensures more rapid training convergence \citep{Wan19}.
\begin{figure*}
    \centering
  \begin{subfigure}{0.9\columnwidth}
    \centering
    \includegraphics[scale=0.32, trim={4.9cm 0 1cm 0}, clip]{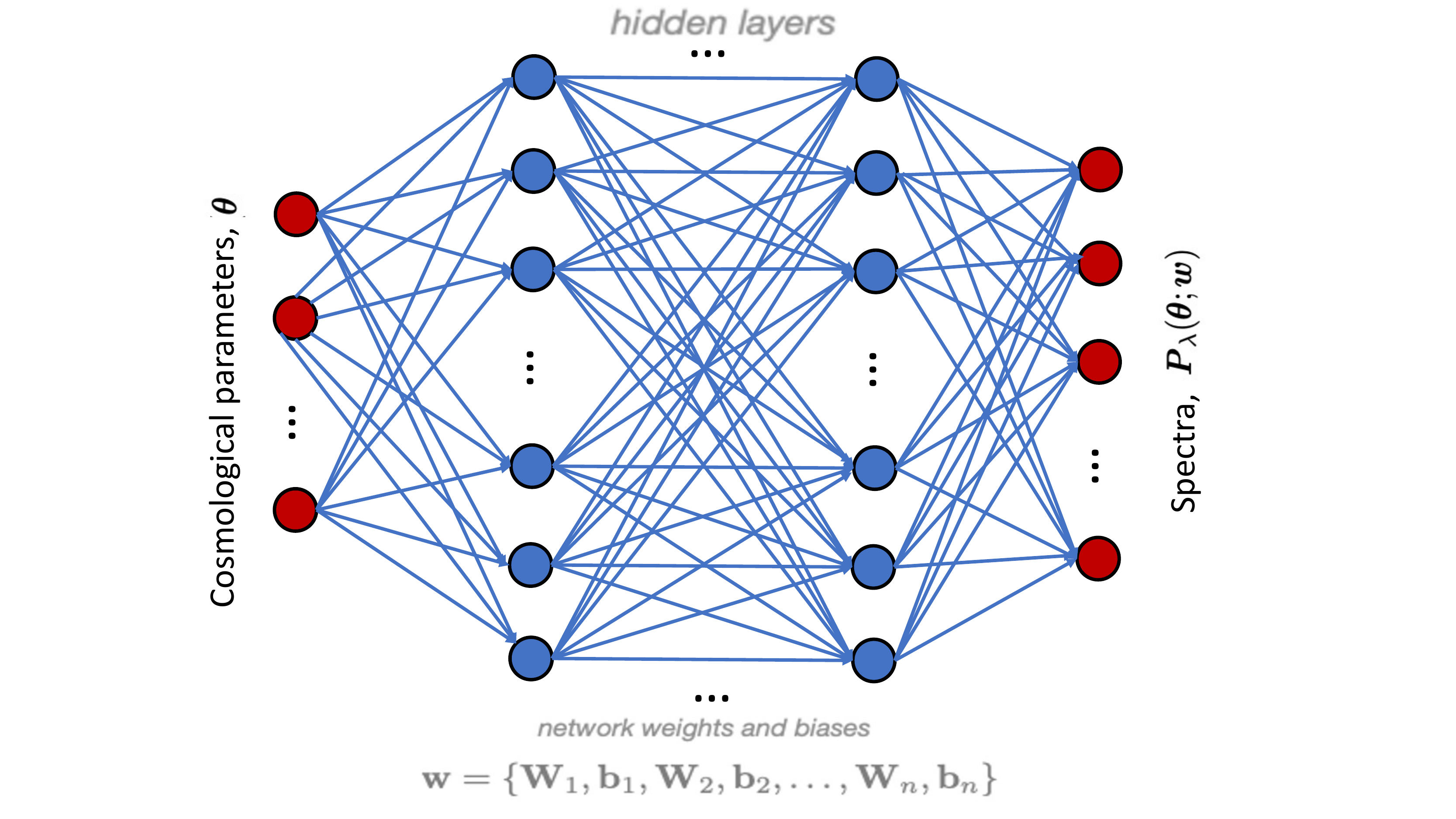}
    \caption[]%
            {}{}
  \end{subfigure}
  \begin{subfigure}{1.15\columnwidth}
    \centering
    \includegraphics[scale=0.32, trim={0.2cm 0 0 0}, clip]{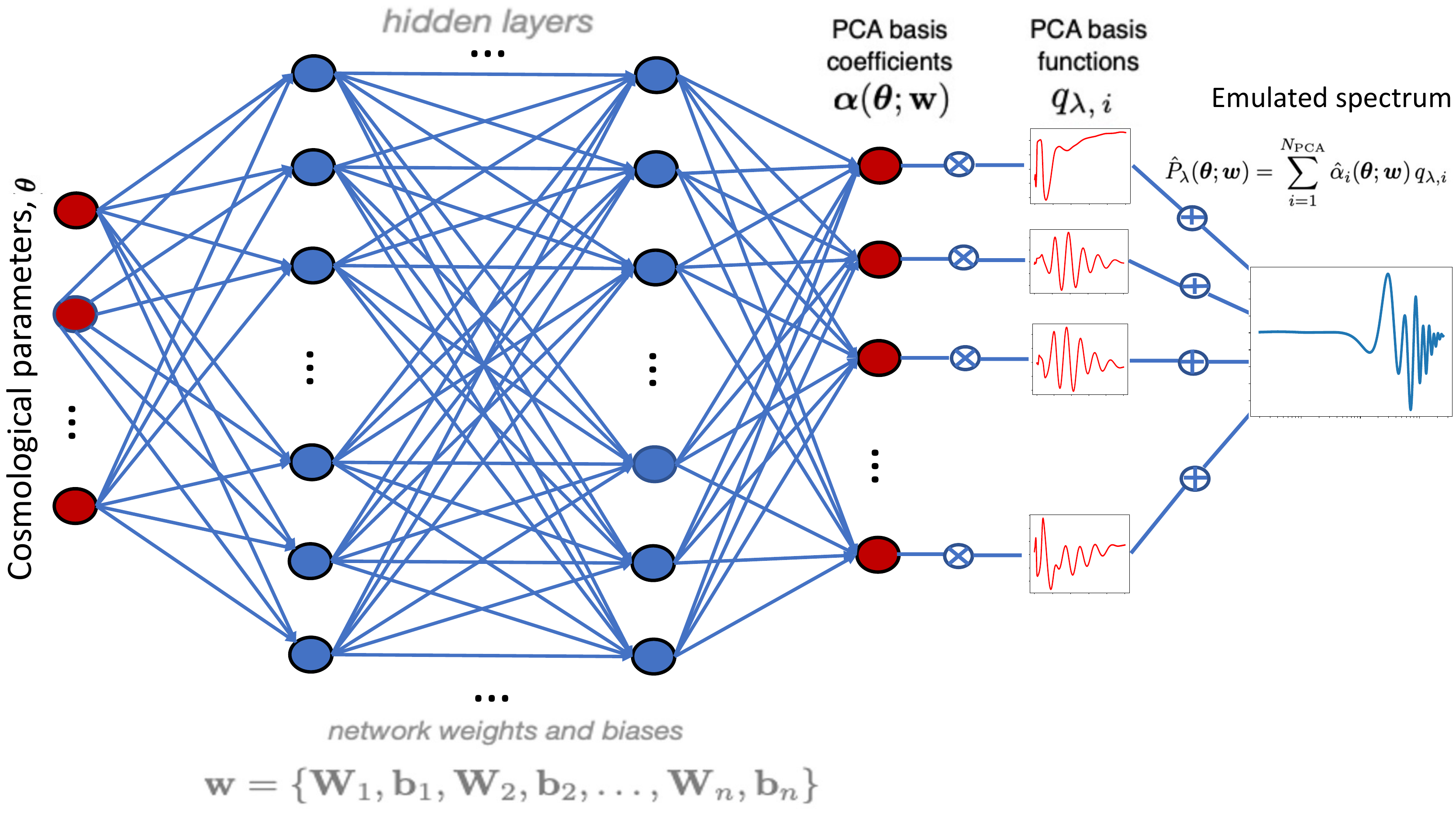}
    \caption[]%
            {}{}
  \end{subfigure}
    \caption{Schematic of the neural network emulation implemented in \textsc{CosmoPower}. A neural network is trained to learn the mapping between cosmological parameters and a) power spectra, b) coefficients in a Principal Component Analysis of the power spectra.}
    \label{fig:emulation}
\end{figure*}
    \item In the second method we train a NN to learn the mapping between cosmological parameters and Principal Components of the power spectra, as shown in the right-hand panel of Fig.~\ref{fig:emulation}. Principal Component Analysis (PCA) is a linear dimensionality reduction technique which performs a singular value decomposition of the input signal and retains the $N_{\mathrm{PCA}}$ components with the highest variance. We perform a PCA decomposition of the spectra in our training dataset, which produces a set of basis functions $q_{\lambda, i}$, with $i \in 1, \dots , N_{\mathrm{PCA}}$. In other words, we assume we can decompose the spectra as:
    \begin{align}
        P_{\lambda}(\boldsymbol{\theta}; \boldsymbol{w}) = \sum_{i=1}^{N_{\mathrm{PCA}}} \alpha_i (\boldsymbol{\theta}; \boldsymbol{w}) q_{\lambda,i} \ ,
    \end{align}
     where the coefficients $\alpha_i$ in the new basis $q_{\lambda, i}$ are the Principal Components. We train a NN to output estimates $\hat{\alpha}_i$ of the Principal Components $\alpha_i$, given cosmological parameters $\boldsymbol{\theta}$ as input. Similarly to the power spectra components in the direct NN case, the PCA components are also standardised.
\end{itemize}
We report the implementation details of the neural networks and PCA in Appendix \ref{app:methods}, including details on the training procedure. We tested both emulation approaches on the cosmological power spectra of interest and found the former to be in general more accurate. Thus, we decided to use it for all power spectra with the exception of the cross temperature-polarisation $\left ( C_{\ell}^{\textrm{TE}} \right )$ and lensing potential $\left ( C_{\ell}^{\phi \phi} \right )$ CMB power spectrum, which were emulated using the second method. The use of the PCA decomposition is indeed particularly convenient when, as in the CMB TE spectrum case, it is not possible to take the logarithm of the training power spectra, due to some negative values. We also observed a better performance of the emulator for the $\phi \phi$ spectrum when applying the PCA compression first; we argue this is due to the smaller values of the logarithmic spectra, which range from $-6$ to $-20$ and might therefore cause numerical issues if fed directly into the NN.

One of the key advantages of \textsc{CosmoPower} is that the emulators are trained on very broad parameter ranges, which we report in Table~\ref{tab:emu_ranges}. The choice of such large parameter ranges is motivated by the desire to provide a tool that is as general as possible (see Sect.~\ref{sec:conclusions} for further comments on this point). 

A common problem in existing emulators is that they are trained to provide good performance on a \textit{fixed} choice of cosmological parameterisation, while Boltzmann solvers maintain the flexibility to choose among different input parameterisations. In addition, Boltzmann solvers allow for derived parameters to be computed \textit{a posteriori}, so that one can explore degeneracies between different parameters without restrictions. For example, one cannot directly sample in both $\mathrm{ln} 10^{10} A_{\mathrm{s}}$ and $\sigma_8$, as these two parameters are related; choosing to sample one or the other may in fact even have some effect on the posterior distribution \citep[see e.g.][]{Joachimi20}. The common strategy when performing cosmological inference is to choose one of these parameters for sampling, and let the Boltzmann solver compute the corresponding other one at each point in the posterior chain. In \textsc{CosmoPower} this is also possible, \textit{without} re-training emulators from scratch. As we show in Sect.~\ref{sec:LSS} and Appendix \ref{app:gp} for the case of $\mathrm{ln} 10^{10} A_s$ and $\sigma_8$, one can choose to sample the former or the latter, and obtain the other one with a GP. We refer the reader to Appendix \ref{app:gp} for details on the implementation of such a GP (see also \citealt{Mootoovaloo20} for a similar application of GPs to obtain derived values of $\sigma_8$).

\section{Large-Scale Structure}\label{sec:LSS}

\subsection{Theory}
Here we briefly summarise the theory underlying two-point statistics analyses of LSS surveys, following a notation similar to that of \citet{Asgari20, Heymans20, Joachimi20}. A flat $\Lambda$CDM model is assumed throughout the paper. Extensions of our emulators to beyond-$\Lambda$CDM cosmologies will be explored in future work (see Sect.~\ref{sec:conclusions} for details).

LSS analyses target the shear and clustering signal of the observed galaxies, including the shear-clustering cross-correlation, in what is typically referred to as `3x2pt' analysis \citep{Joachimi10}. Angular power spectra of shear and clustering statistics can be expressed as integrals of the matter power spectrum $P_{\delta \delta}(k, z)$ along the line of sight, weighted by kernel functions:
\begin{align}\label{eq:limber}
C_{ij}^{\rm ab}(\ell) = \int^{\chi_{\rm H}}_0 \!\!\! \mathrm{d} \chi\;
\frac{W^{\rm a}_{i} (\chi)\; W^{\rm b}_{ j} (\chi)}{\chi^2}\; P_{\delta \delta} \left( \frac{\ell+1/2}{\chi}, z \right) \ ,
\end{align}
where the indices $\{ {\rm a,b} \}$ can be assigned the labels $\{\gamma, \mathrm{I}, \mathrm{n}\}$, denoting contributions from cosmic shear, galaxy intrinsic alignment and galaxy positions, respectively. The integral in Eq.~(\ref{eq:limber}) is performed along the line of sight up to the Hubble radius $\chi_{\rm{H}} = c/H_0$, with $c$ the speed of light and $H_0$ the Hubble constant. $\chi$ denotes the comoving distance, which is a function of the redshift $z$ (a dependence implicitly assumed in Eq.~\ref{eq:limber} for ease of notation). The Limber projection \citep{Kaiser92, LoVerde08} connects the wavenumber $k$ and redshift $z$ at which the matter power spectrum $P_{\delta \delta}(k, z)$ is evaluated in that integral, so that given a multipole $\ell$ and a redshift $z$ (corresponding to a comoving distance $\chi$) the matter power spectrum is evaluated at wavenumber $k = (\ell + 1/2) / \chi(z)$. We note that, while we will restrict ourselves to Limber-approximated spectra in this paper, the emulation framework in \textsc{CosmoPower} can be equally applied to accelerate the computation of non-Limber projected spectra, which we plan to investigate in future work (see Sect.~\ref{sec:conclusions} for a discussion).

The weighting functions $W$ can be written as 
\begin{align}\label{eq:kernels}
W^{\gamma}_{i} (\chi) &= \frac{3 \, H_0^2 \, \Omega_{\mathrm{m}}}{2\, c^2}
\frac{\chi}{a} \int_{\chi}^{\chi_{\mathrm{H}}} \mathrm{d} \chi'\; n_{i}(\chi')\; \frac{\chi' - \chi}{\chi'}\; \ , \\
W^{\rm I}_{i} (\chi) &= - A_{\rm IA}\; \left( \frac{1+z}{1+z_{\rm pivot}} \right)^{\eta_{\rm IA}}  \frac{C_1 \, \rho_{\rm cr} \, \Omega_{\rm m}}{D(\chi)}\; n_{i}(\chi)\; \ , \\
W^{\rm n}_{i} (\chi) &= b_i \, n_{i}(\chi) \ ,
\end{align}
where $\Omega_{\mathrm{m}}$ is the total matter density parameter, $a$ is the scale factor, $n_i(\chi)$ denotes the redshift distribution for redshift bin $i$, $D(\chi)$ is the linear growth factor, $\rho_{\mathrm{cr}}$ is the critical density, $C_1$ is a constant, $z_{\mathrm{pivot}}$ is an arbitrary redshift set to 0.3, while $A_{\mathrm{IA}}$ and $\eta_{\mathrm{IA}}$ are two free parameters, allowing for a scaling in amplitude and redshift, respectively, of the intrinsic alignment contribution. The linear galaxy bias coefficients $b_i$ are also left free to vary in the inference pipeline.

The observed angular two-point statistics of galaxy ellipticities $\epsilon$ for tomographic redshift bins $i$ and $j$, as a function of the angular multipole $\ell$, can be written as
\begin{align}\label{eq:cl_wl}
    C^{\epsilon \epsilon}_{ij}(\ell) = C^{\gamma \gamma}_{ij}(\ell) + C^{\gamma \mathrm{I}}_{ij}(\ell) + C^{\mathrm{I}\gamma}_{ij}(\ell) + C^{\mathrm{II}}_{ij}(\ell) \ ,
\end{align}
i.e. as a sum of a pure cosmic shear contribution and contaminants resulting from the intrinsic alignment of galaxies, which also affects the angular power spectrum of the cross-correlation between galaxy ellipticities $\epsilon$ and positions $n$ (usually referred to as `galaxy-galaxy lensing'):
\begin{align}\label{eq:cl_ggl}
    C^{\mathrm{n} \epsilon}_{ij}(\ell) &= C^{\mathrm{n} \gamma}_{ij}(\ell) + C^{n \mathrm{I}}_{ij}(\ell) \ ,
\end{align}
while the galaxy clustering power spectrum $C^{\mathrm{n n}}_{ij}(\ell)$ is not affected by intrinsic alignments and, assuming no \textit{magnification bias} \citep{Duncan13, von_Wietersheim_Kramsta21}, can be explicitly written as 
\begin{align}\label{eq:cl_gg}
    C^{\mathrm{n n}}_{ij}(\ell) = b_i b_j \int_0^{\chi_\mathrm{H}} \mathrm{d} \chi \, \frac{n_{i}(\chi) \, n_{j}(\chi)}{\chi^2} \,  P_{\delta \delta} \left( \frac{\ell+1/2}{\chi}, z \right). 
\end{align}
Two-point statistics measured by LSS surveys are typically linear transformations of the theoretical power spectra given by Eqs.~\ref{eq:cl_wl}, \ref{eq:cl_ggl}, \ref{eq:cl_gg}, such as band-power estimates \citep{Schneider02, vanUitert18}. We refer the reader to \citet{Asgari20} for a review of different measured two-point statistics.

\begin{table}
  \centering 
  \begin{tabular}{c|c|c}
    \textbf{Parameter}               &  \textbf{LSS emulator range} & \textbf{CMB emulator range} \\
    \hline
    \hline
    $\omega_{\mathrm{b}}$   & [0.01875, 0.02625] & [0.005, 0.04]\\
    \hline
    $\omega_{\mathrm{cdm}}$ & [0.05, 0.255]      & [0.001, 0.99]\\
    \hline
    $h$                     & [0.64, 0.82]       & [0.2, 1.0]\\
    \hline
    $\tau_{\mathrm{reio}}$                  & fixed              & [0.01, 0.8]\\
    \hline
    $n_s$                   & [0.84, 1.1]        & [0.7, 1.3]\\
    \hline
    $\mathrm{ln}10^{10}A_s$ & [1.61, 3.91]       & [1.61, 5]\\
    \hline
    $c_{\mathrm{min}}$      & [2, 4]           & fixed \\
    \hline
    $\eta_{0}$              & [0.5, 1]          & fixed \\
    \hline
    \hline
    \end{tabular}
   \caption{Ranges of validity of our emulators. The uniform prior distributions that we use at inference time to test our emulators share the same ranges. The LSS and CMB ranges are also the prior ranges assumed by the KiDS-1000 and Planck 2018 analysis, respectively. Parameters denoted as \textit{fixed} in the LSS or CMB column are not considered as inputs to the neural networks that emulate the relevant power spectra, since the power spectra dependence on those parameters is negligible.}
  \label{tab:emu_ranges}
\end{table}

\subsection{Emulating the matter power spectrum}
\begin{figure*} 
  \begin{subfigure}{\columnwidth}
    \centering
    \includegraphics[scale=0.55]{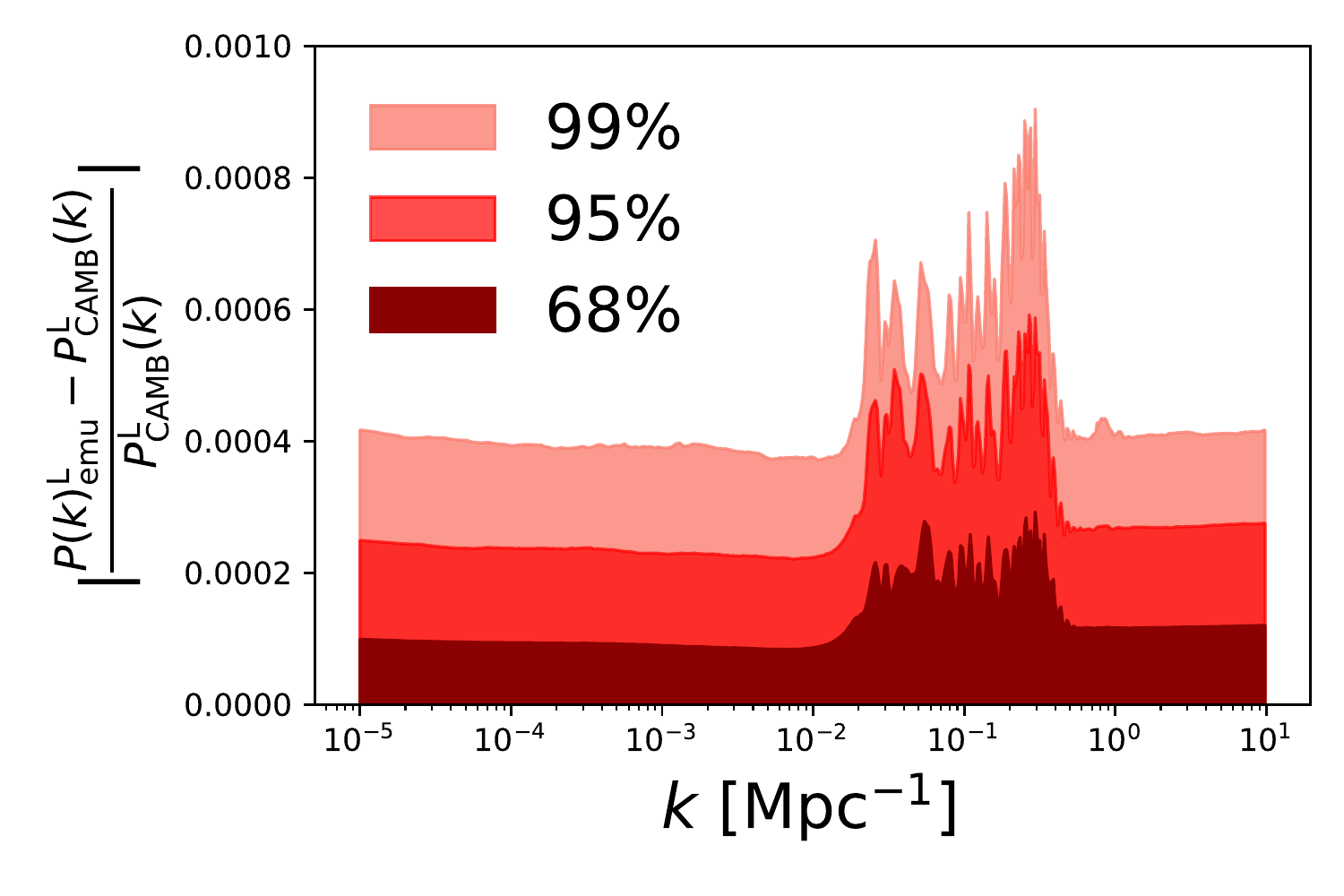}
    \caption[]%
            {}{}
  \end{subfigure}
  \begin{subfigure}{\columnwidth}
    \centering
    \includegraphics[scale=0.55]{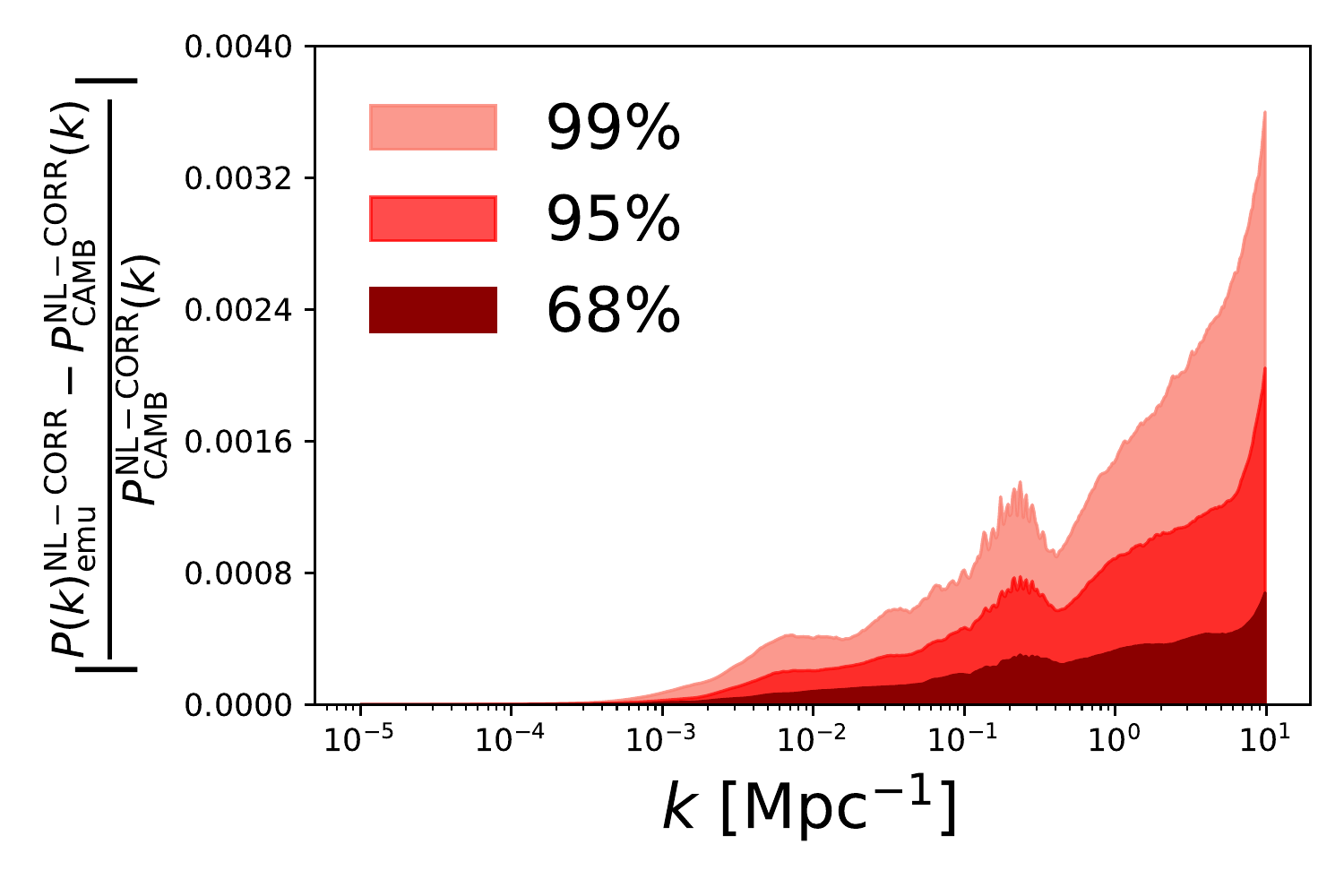}
    \caption[]%
            {}
  \end{subfigure}
  \caption{Matter power spectrum emulation accuracy for a) the linear power spectrum and b) the nonlinear correction, as measured on the $\sim 2 \cdot 10^4$ spectra of the testing set for our emulators. \textit{Dark red}, \textit{red} and \textit{salmon} areas enclose the 68, 95 and 99 percentiles of the fractional absolute emulator error, respectively, as a function of wavenumber $k$. By construction, the redshift $z$ is an input parameter for the emulators. Therefore, these percentile curves are computed with spectra evaluated at redshifts $z \in [0, 5]$, i.e. spanning the whole range of validity of our emulators.}
  \label{fig:acc_mnu0}
\end{figure*}

Eq.~(\ref{eq:limber}) clearly indicates that the prediction of the matter power spectrum $P(k, z)$ is central in the computation of the two-point statistics of cosmic shear, galaxy-galaxy lensing and galaxy clustering. Boltzmann codes can perform practically exact computations (up to numerical accuracy) of the matter power spectrum predicted by the linear theory of gravitational instability, which we denote as $P_{\delta \delta}^{\rm{L}} (k, z)$. As nonlinearities become more important, linear theory breaks down. We will write the full, nonlinear spectrum as the product of the linear one $P_{\delta \delta}^{\rm{L}} (k, z)$ and a nonlinear correction, which we label $P_{\delta \delta}^{\rm{NL-CORR}} (k, z)$:
\begin{align}
    P_{\delta \delta} (k, z) = P_{\delta \delta}^{\rm{L}} (k, z) \, P_{\delta \delta}^{\rm{NL-CORR}} (k, z) \ .
\end{align}

Nonlinear corrections become important at small scales, corresponding to large wavenumbers $k$. Their modelling as a function of cosmological parameters is uncertain and further complicated by baryonic effects, whose impact on the nonlinear matter power spectrum induces important, yet not fully understood modifications to the nonlinear power on small scales \citep[see e.g.][for a review]{Chisari19}. In recent years, the \textsc{HMcode} software developed by \citet{Mead15, Mead16, Mead21} has found widespread use in LSS analyses, as a way to predict the nonlinear power spectrum while taking into account baryonic effects. \textsc{HMcode} is a modified halo model which includes baryonic contributions as opposed to, for example, the fitting function \textsc{HaloFit} \citep{Smith03, Takahashi12, Bird12}. In this work we consider the latest version of the \textsc{HMcode} software \citep{Mead21}. We focus on two of its free parameters, $c_{\mathrm{min}}$ and $\eta_0$, denoting the minimum halo concentration and the halo bloating, respectively. We train one emulator for the linear power spectrum $P_{\delta \delta}^{\mathrm{L}} (k, z)$ and one for the nonlinear correction $P_{\delta \delta}^{\mathrm{NL-CORR}} (k, z)$. We also experimented emulating directly the full power spectrum, but noticed increased performance when separating the linear contribution from the nonlinear correction. We observe this split helps the emulator isolate and learn more efficiently the effect of the nonlinear parameters $c_{\mathrm{min}}$ and $\eta_0$. For both linear and nonlinear contribution, we emulate the power spectrum at 420 $k$ values in the range $[10^{-5}, 10] \, \mathrm{Mpc}^{-1}$. The redshift $z$ is varied over the range $[0, 5]$ and treated as an additional input parameter for the emulator. We use $\sim 1.8 \cdot 10^5$ spectra for our training set and leave $\sim 2 \cdot 10^4$ spectra for our testing set. We verified that this fraction of parameter samples randomly selected for testing purposes from the training set still represents a homogeneous and uniformly distributed sample for each parameter.

The left- and right-hand panels of Fig.~\ref{fig:acc_mnu0} report the emulation accuracy on the testing set for the linear power spectrum and nonlinear correction, respectively. We use percentile plots to show the statistical behaviour of the emulator accuracy throughout the testing set, as a function of the wavenumber $k$. For the linear power, 99\% of the testing power spectra are emulated with differences smaller than 0.1\% of their real value across the entire wavenumber range considered. For the nonlinear correction, 99\% of the spectra are emulated with less than 0.4\% error. As expected, the percentage difference in the nonlinear correction is typically about an order of magnitude larger than the linear one, thus dominating the total error. We can see how the error on the nonlinear correction increases for $k \gtrsim 1 \, \mathrm{Mpc}^{-1}$, i.e. on highly non-linear scales. This was expected, as numerical computations performed by Boltzmann codes at these scales are more uncertain in the first place. One way to ensure increased numerical stability in these computations is to ask for a maximum $k$ value $k_{\mathrm{max}}$ at which nonlinearities are computed that is well above the actual maximum $k$ at which the power spectrum is required. In our analysis we set $k_{\mathrm{max}}=100 \, \mathrm{Mpc}^{-1}$, while we only use the matter power spectrum up to $k = 10 \, \mathrm{Mpc}^{-1}$. 

These accuracy plots already show excellent performance of the emulators in obtaining high-fidelity predictions for the matter power spectrum. However, a full inference analysis is required to thoroughly test our emulators. As further discussed in Sect.~\ref{sec:conclusions}, we remark here that this is the only way to completely test the validity of an emulation approach such as the one developed in \textsc{CosmoPower}. Crucially, different accuracy emulation levels are required for different types of analyses for which the emulators are being developed. \textsc{CosmoPower} is a tool designed to be adequate for Stage IV surveys: as such, we need to test the performance of \textsc{CosmoPower} on a simulated Stage IV inference pipeline. This is what we show in Sect.~\ref{sec:euclid} with the simulated analysis of a \textit{Euclid}-like survey.

Before showing those results, we present in Sect.~\ref{sec:kidsxgama} an application to a Stage III analysis, namely a joint weak lensing and galaxy clustering analysis from 450 deg$^2$ of the KiDS survey. We refer the reader to Appendix \ref{app:kids1000} for another, more recent Stage III application, namely a cosmic shear analysis of 1000 deg$^2$ from the KiDS survey. The goal of showing these Stage III results is to demonstrate that while \textsc{CosmoPower} is explicitly designed to aid inference analyses from Stage IV surveys, it can clearly benefit those already ongoing from Stage III experiments too. In the latter, \textsc{CosmoPower} can indeed provide contours in a matter of a few minutes without any Graphic Processing Unit acceleration (see Sect.~\ref{sec:conclusions} for a discussion on this point). In addition, in showing results for Stage III analyses we would like to emphasise the `train-once-use-repeatedly' nature of \textsc{CosmoPower}. Our emulators are designed to be trained only once, so that the end users do not need to re-train any of the models, as long as the emulators are employed assuming the cosmological model and range indicated in Table~\ref{tab:emu_ranges}. For example, in the KiDS and \textit{Euclid}-like analyses shown in Sect.~\ref{sec:kidsxgama} and \ref{sec:euclid}, the emulators for the linear power spectrum and nonlinear correction are the \textit{same} for both analyses. 

\subsection{Validation on the KiDS-450+GAMA 3x2pt likelihood}\label{sec:kidsxgama}
\begin{table}
  \centering 
  \begin{tabular}{c|c}
    \textbf{Parameter}               &  \textbf{Prior range} \\
    \hline
    \hline
    $\omega_{\mathrm{b}}$   & [0.01875, 0.02625] \\
    \hline
    $\omega_{\mathrm{cdm}}$ & [0.05, 0.255]      \\
    \hline
    $h$                     & [0.64, 0.82]       \\
    \hline
    $n_s$                   & [0.84, 1.1]        \\
    \hline
    $\mathrm{ln}10^{10}A_s$ & [1.61, 3.91]       \\
    \hline
    $c_{\mathrm{min}}$      & [2, 4]           \\
    \hline
    $A_{\mathrm{IA}}$       & [-6, 6]          \\
    \hline
    b$_{z_1}$               & [0.1, 5]          \\
    \hline
    b$_{z_2}$               & [0.1, 5]          \\
    \hline
    \hline
    \end{tabular}
   \caption{Prior ranges for the 3x2pt analysis of the KiDS-450 and GAMA surveys. Prior distributions are all taken to be uniform across these ranges. For the cosmological parameters and the baryonic feedback parameter $c_{\mathrm{min}}$ the prior range corresponds to the range of validity of our emulators (cf. Table \ref{tab:emu_ranges}).}
  \label{tab:prior_kxg}
\end{table}

\begin{figure*}
    \centering
    \includegraphics[width=\textwidth]{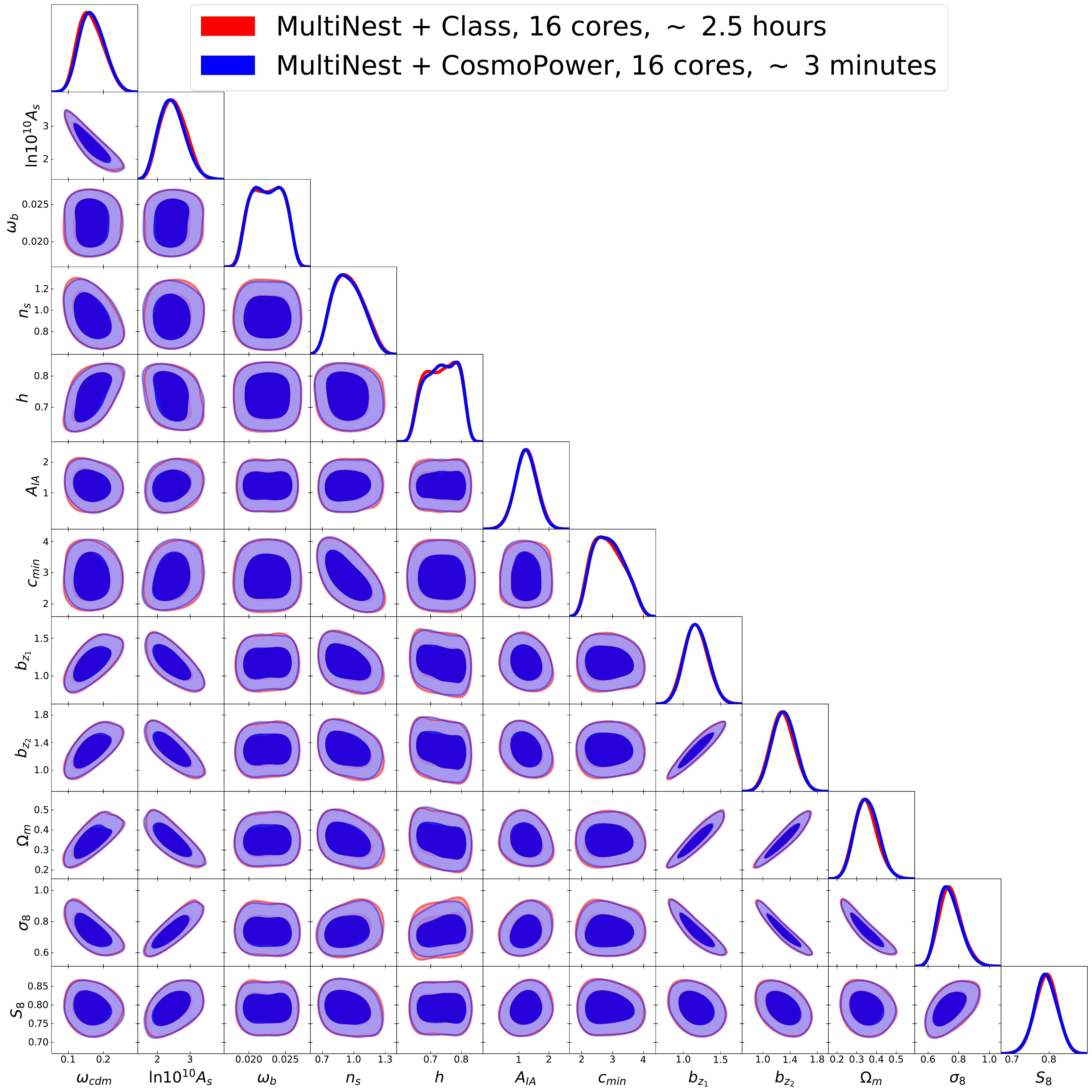}
    \caption{Contours  for the 3x2pt analysis of the KiDS-450 and GAMA surveys. In \textit{red} contours obtained with \textsc{Class}, in \textit{blue} those obtained with \textsc{CosmoPower}.}
    \label{fig:kidsxgama_contours}
\end{figure*}

\citet{vanUitert18} performed a 3x2pt analysis of power spectra from 450 $\mathrm{deg}^2$ of the KiDS survey (KiDS-450) with overlapping galaxy clustering spectroscopic measurements from the Galaxy And Mass Assembly survey \citep[GAMA,][]{Driver09, Driver11, Liske15}. Here, we repeat their analysis using an inference pipeline entirely embedded within the cosmological sampler \textsc{MontePython} \citep{Brinckmann18}. This pipeline is an adaptation to $\Lambda$CDM of the likelihood developed in \citet{SpurioMancini19}\footnote{available at \url{https://github.com/alessiospuriomancini/KiDSHorndeski}} for the same survey configuration, but for a Horndeski gravity scenario. In \citet{SpurioMancini19} that likelihood, run in $\Lambda$CDM, was already shown to provide excellent agreement with the fiducial pipeline developed by \citet{vanUitert18} for the sampler \textsc{CosmoMC}. In addition to the data vector, the redshift distributions and analytical covariance matrix are also exactly the same as those used in \citet{vanUitert18} and \citet{SpurioMancini19}. 

The modelling of the power spectra for each cosmological probe follows the same prescription as in \citet{vanUitert18, SpurioMancini19} and described in Eqs.~\ref{eq:cl_wl}, \ref{eq:cl_ggl}, \ref{eq:cl_gg}. In particular, we consider one intrinsic alignment amplitude $A_{\mathrm{IA}}$, we set $\eta_{\mathrm{IA}}=0$ and we include one linear galaxy bias coefficient $b_{z_i}$ for each of the two spectroscopic bins of the GAMA survey. The prior ranges are reported in Table \ref{tab:prior_kxg}. They are the same ones used in \citet{vanUitert18, SpurioMancini19}, except for those cosmological parameters ($\omega_{\mathrm{cdm}}, \omega_{\mathrm{b}}, n_{\mathrm{s}}$), whose prior range was restricted in the more recent KiDS-1000 analysis \citep{Asgari20, Heymans20, Joachimi20}. As explained in \citet{Joachimi20}, the KiDS collaboration decided to restrict those prior ranges in their analyses, to make sure that the parameters are more physically motivated; we adhere to this choice. For the cosmological parameters in Table \ref{tab:prior_kxg} and the baryonic feedback parameter $c_{\mathrm{min}}$, the prior ranges coincide with the range of validity of our emulators (cf. Table \ref{tab:emu_ranges}). The halo bloating parameter $\eta_0$ is fixed to $\eta_0 = 0.98 - 0.12 c_{\mathrm{min}}$, as implemented in \textsc{Class} following \citet{Mead15}.

The recent KiDS-1000 analyses sample the cosmological parameter $S_8 = \sigma_8 \sqrt{\Omega_{\mathrm{m}}/0.3}$, instead of $\mathrm{ln} 10^{10} A_s$ used in \citet{vanUitert18} and other past KiDS analyses \citep{Hildebrandt16, Kohlinger17, Hildebrandt20}. Here, we stick to the choice of $\mathrm{ln} 10^{10} A_s$ as in \citet{vanUitert18}, for a more direct comparison. In Appendix \ref{app:kids1000}, instead, we show results for KiDS-1000 with $S_8$ among the sampled parameters. In all cases, we always show plots for all of $\mathrm{ln} 10^{10} A_s$, $\sigma_8$ and $S_8$: the parameters that have not been used directly in sampling have been obtained as derived parameters with a postprocessing GP, as explained in Sect.~\ref{sec:methods} and Appendix \ref{app:gp}.

Fig.~\ref{fig:kidsxgama_contours} shows the comparison between contour plots obtained sourcing power spectra from \textsc{Class} and those obtained running \textsc{CosmoPower} to replace the Boltzmann code. In both analyses, we sample the posterior distribution with the \textsc{MultiNest} algorithm \citep{Feroz09}, as implemented in the Python wrapper \textsc{PyMultiNest} \citep{Buchner14}. We run the pipelines with parallelisation across 16 Intel Xeon E5640 cores. The posteriors with our emulator are obtained in under 3 minutes, compared with the $\sim$ 2.5 hours required when using \textsc{Class}. This produces a speed-up factor of approximately 50. 

We also compare the values of the log-evidence $\mathrm{log} \, \mathcal{Z}$ obtained with \textsc{MultiNest} in the two runs, and find $\mathrm{log} \, \mathcal{Z} = -73.77 \pm 0.19$ and $\mathrm{log} \, \mathcal{Z} = -73.79 \pm 0.19$ for the run sourcing power spectra from \textsc{Class} and \textsc{CosmoPower}, respectively. The good agreement between these evidence values is another, arguably stronger indicator of the accuracy of our emulators. In order to get an accurate estimate of this quantity one needs indeed accurate values of the posterior distribution across the whole prior range, while for parameter estimation it may be sufficient to get accurate estimates around the peak of the distribution.

We also applied \textsc{CosmoPower} to a more recent KiDS analysis, namely that considering cosmic shear band powers from 1000 deg$^2$ of the survey, as presented in \citet{Asgari20}. We use this analysis as another test case for the performance of our emulator, finding similar speed-up factors provided by replacing the Boltzmann code \textsc{Class} with \textsc{CosmoPower}. We refer the reader to Appendix \ref{app:kids1000} for details and plots for this analysis.

\subsection{Validation on the Euclid-like cosmic shear likelihood}\label{sec:euclid}

\begin{table}
  \centering 
  \begin{tabular}{c|c|c}
    \textbf{Parameter}               &  \textbf{Prior range}  & \textbf{Fiducial value} \\
    \hline
    \hline
    $\omega_{\mathrm{b}}$            & [0.01875, 0.02625]     & 0.02242\\
    \hline
    $\omega_{\mathrm{cdm}}$          & [0.05, 0.255]          & 0.11933\\
    \hline
    $h$                              & [0.64, 0.82]           & 0.6766\\
    \hline
    $n_s$                            & [0.84, 1.1]            & 0.9665\\
    \hline
    $\mathrm{ln}10^{10}A_s$          & [1.61, 3.91]           & 3.047\\
    \hline
    $c_{\mathrm{min}}$               & [2, 4]                 & 2.6\\
    \hline
    $\eta_0$                         & [0.5, 1]               & 0.7\\
    \hline
    $A_{\mathrm{IA}}$                & [-6, 6]                & 0.8\\
    \hline
    $\eta_{\mathrm{IA}}$             & [-6, 6]                & 0\\
    \hline
    $D_{z_i}, \quad i = 1 \dots 10$  & $\mathcal{N}(0, 10^{-4})$  & 0\\
    \hline
    \hline
    \end{tabular}
   \caption{Prior ranges and fiducial values of the cosmological parameters for the simulated \textit{Euclid}-like cosmic shear analysis. Prior distributions are all taken to be uniform across these ranges, except for the redshift mean shifts $D_{z_i}$ which have a Gaussian prior with mean 0 and standard deviation $10^{-4}$. For the cosmological parameters and the baryonic feedback parameters $c_{\mathrm{min}}, \eta_0$ the prior range corresponds to the range of validity of our emulators (cf. Table \ref{tab:emu_ranges}).}
  \label{tab:prior_euclid}
\end{table}

\begin{figure*}
    \centering
    \includegraphics[width=1.05\textwidth, height=22cm]{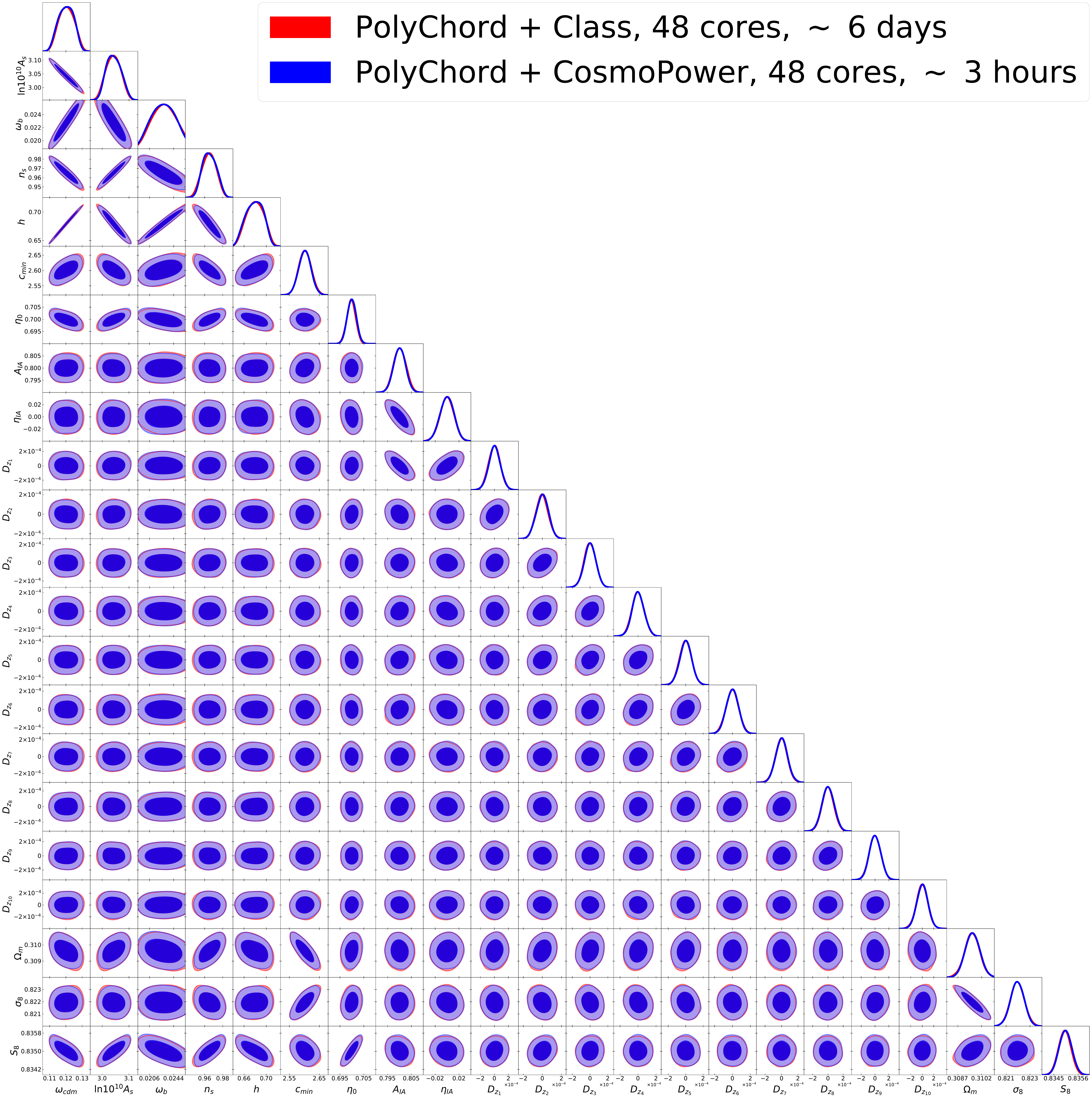}
    \caption{Contours for a simulated cosmic shear analysis of a \textit{Euclid}-like Stage IV surveys. In \textit{red} contours obtained with \textsc{Class}, in \textit{blue} those obtained with \textsc{CosmoPower}.}
    \label{fig:euclid_contours}
\end{figure*}

We re-use the same emulator employed in Sect.~\ref{sec:kidsxgama} to perform a full inference analysis of a simulated cosmic shear survey for a typical Stage IV \textit{Euclid}-like configuration. The prior ranges are reported in Table \ref{tab:prior_euclid}. For the cosmological and baryonic feedback parameters they correspond to the validity ranges of our emulators, reported in Table \ref{tab:emu_ranges}. The fiducial values of the parameters used to calculate the mock data vector are also reported in Table \ref{tab:prior_euclid}. 10 tomographic bins are equipopolated with galaxies following a distribution \citep{Smail95, Joachimi10, Laureijs11}:
\begin{align}
    n(z) \propto z^2 \exp \{ -(z/z_0)^{3/2} \}
\end{align}
with $z_0=0.64$. For the covariance matrix, we use a simple analytical Gaussian computation following \citet{Tutusaus20}, with a sky coverage $f_{\mathrm{sky}}=0.3$, a galaxy number density $n = 30 \ \mathrm{arcmin}^{-2}$ and an ellipticity dispersion $\sigma_{\epsilon}=0.3$.

We implement the likelihood in the cosmological sampler \textsc{Cobaya} \citep{Torrado19, Torrado21} and use the \textsc{Polychord} algorithm \citep{Handley15a, Handley15b} to sample the posterior distribution. We run the pipelines with parallelisation across 48 Intel Xeon E5640 cores and obtain a speed-up of approximately 50. Fig.~\ref{fig:euclid_contours} shows the excellent agreement between contours obtained with \textsc{Class} and \textsc{CosmoPower}. The log-evidence $\mathrm{log} \, \mathcal{Z} = -45.92 \pm 0.33$ computed for the run with spectra sourced from \textsc{Class} also compares favourably with that of the emulator run, $\mathrm{log} \, \mathcal{Z} = -45.99 \pm 0.34$.

\section{Cosmic Microwave Background}
\label{sec:CMB}
\subsection{Emulating CMB temperature, polarisation and lensing power spectra}
The three main probes in the analysis of the CMB are the temperature power spectrum ($C_{\ell}^{\textrm{TT}}$), the polarisation power spectrum ($C_{\ell}^{\textrm{EE}}$), and the temperature-polarisation cross power spectrum ($C_{\ell}^{\textrm{TE}}$). We use \textsc{Camb} to produce the training set, which consists of $\sim 5 \cdot 10^5$ power spectra for each probe, with their associated parameters. The parameters are sampled from the range indicated in Table~\ref{tab:emu_ranges} using Latin Hypercube Sampling. We verified that the specific method chosen for sampling parameter space is comparatively less important than the number of training points. We obtained essentially the same accuracy results in our analysis using alternatives to Latin Hypercube Sampling such as Orthogonal Sampling \citep{Owen92} or even a simple uniform sampling for each parameter. This is due to the fact that differences between different sampling schemes tend to become negligible as the number of parameters becomes large, as it is in our case.

To emulate the CMB spectra, we consider both methods described in Sect.~\ref{sec:methods}. We use the direct NN mapping for $C_{\ell}^{\textrm{TT}}$ and $C_{\ell}^{\textrm{EE}}$, while we  found that the $C_{\ell}^{\textrm{TE}}$ emulation is better performed by the second method, i.e. PCA compression followed by a NN, due to its zero-crossing values. To show the flexibility of our approach, we also train a PCA plus NN emulator on the lensing potential power spectrum $C_{\ell}^{\phi \phi}$; note, however, that this probe is not included in the likelihood analysis performed in the next section.

The accuracy of the emulators over the $\ell$ range is measured with respect to the instrumental noise given by the upcoming Simons Observatory \citep{Ade19} combined with cosmic variance. In particular, we calculate the emulation error as:
\begin{align}
\label{eq:cl_cmb_error}
\frac{| C_{\ell, \textrm{emulated}}^{\{\textrm{TT}, \textrm{EE}, \textrm{TE}, \phi \phi\}} - C_{\ell, \textrm{true}}^{\{\textrm{TT}, \textrm{EE}, \textrm{TE},  \phi \phi\}}|} {\sigma_{\ell, \textrm{CMB}}^{\{\textrm{TT}, \textrm{EE}, \textrm{TE}, \phi \phi\}}} \ ,
\end{align}
where 
\begin{align}
\label{eq:cl_cmb_error_nonte}
\sigma_{\ell, \textrm{CMB}}^{\{\textrm{TT}, \textrm{EE}, \phi \phi\}} = \sqrt{\frac{2}{f_{\textrm{sky}}(2\ell+1)}} \left( C_{\ell, \textrm{true}}^{\{\textrm{TT}, \textrm{EE}, \phi \phi\}} + N_{\ell}^{\{\textrm{TT}, \textrm{EE}, \phi \phi\}} \right) \ , 
\end{align}
\begin{align}
\label{eq:cl_cmb_error_te}
\sigma_{\ell, \textrm{CMB}}^{\textrm{TE}} = & \sqrt{\frac{1}{f_{\textrm{sky}}(2\ell+1)}} \nonumber \\  & \times \sqrt{C_{\ell, \textrm{true}}^{\textrm{TE}} C_{\ell, \textrm{true}}^{\textrm{TE}} + \left( C_{\ell, \textrm{true}}^{\textrm{TT}} + N_{\ell}^{\textrm{TT}} \right) \left( C_{\ell, \textrm{true}}^{\textrm{EE}} + N_{\ell}^{\textrm{EE}}\right)} \ ,
\end{align}
$f_{\textrm{sky}} = 0.4$, and $ N_{\ell}^{\{\textrm{TT}, \textrm{EE}, \textrm{TE}, \phi \phi\}}$ refers to the Simons Observatory goal noise curves due to instrumental and atmospheric effects \citep{Ade19}\footnote{\url{https://github.com/simonsobs/so_noise_models}}. 

\begin{figure*}
        \centering
        \begin{subfigure}[b]{0.475\textwidth}
            \centering
            \includegraphics[width=\textwidth]{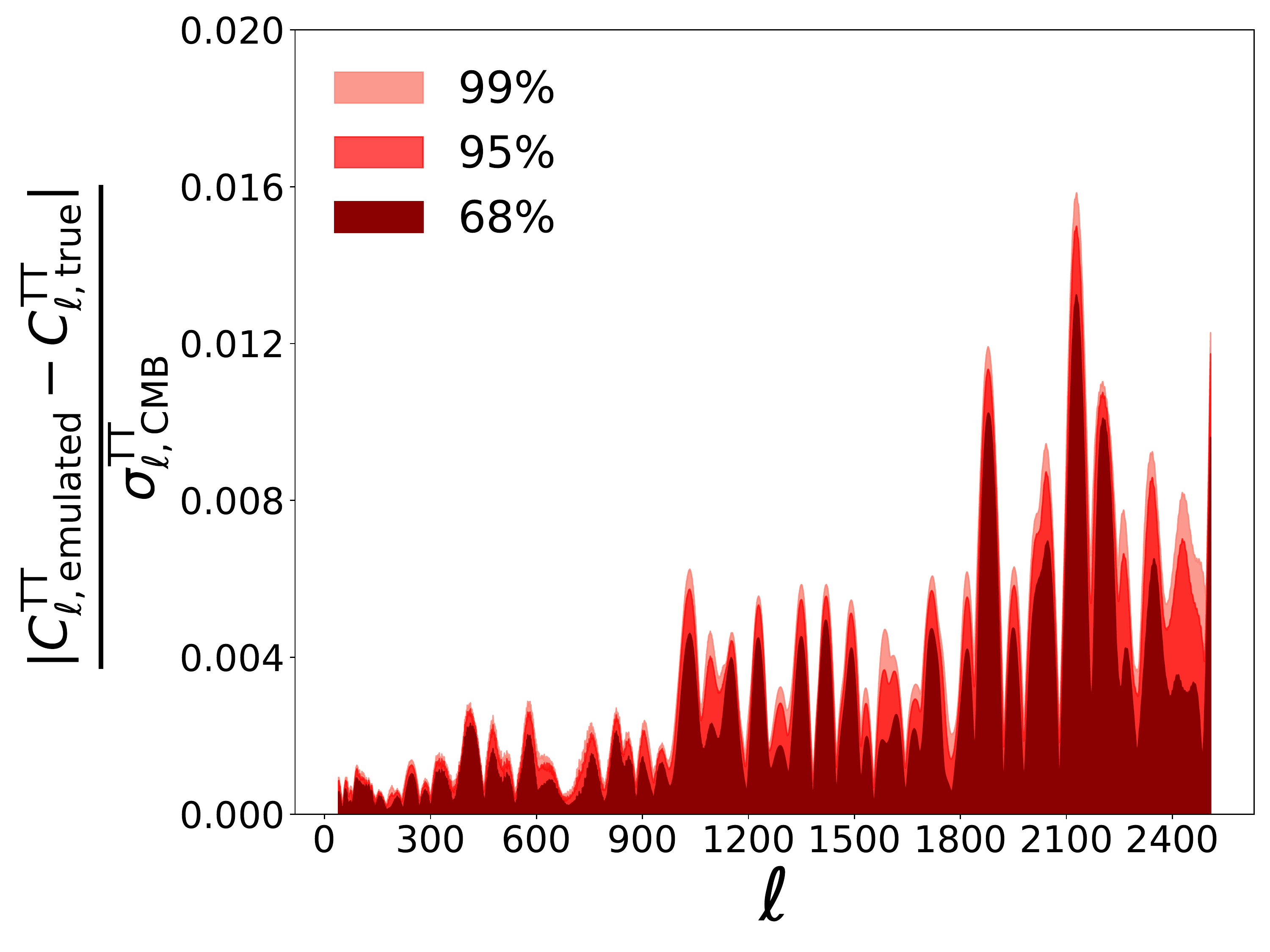}
            \caption[]%
            {{\small $C_{\ell}^{\textrm{TT}}$ }}    
            \label{fig:planck_tt_emu_narrow}
        \end{subfigure}
        \hfill
        \begin{subfigure}[b]{0.475\textwidth}  
            \centering 
            \includegraphics[width=\textwidth]{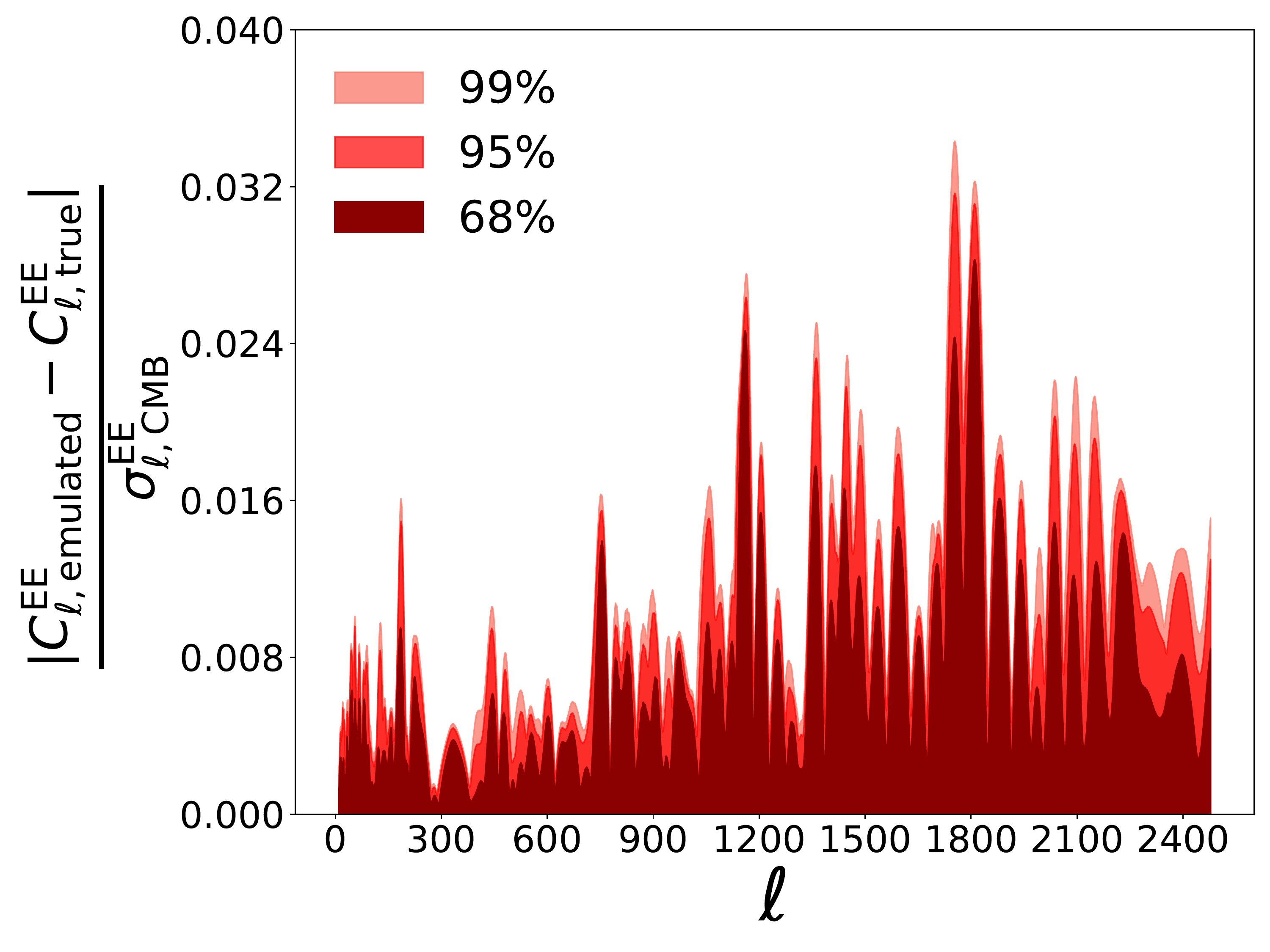}
            \caption[]%
            {{\small $C_{\ell}^{\textrm{EE}}$}}    
            \label{fig:planck_ee_emu_narrow}
        \end{subfigure}
        \vskip\baselineskip
        \begin{subfigure}[b]{0.475\textwidth}   
            \centering 
            \includegraphics[width=\textwidth]{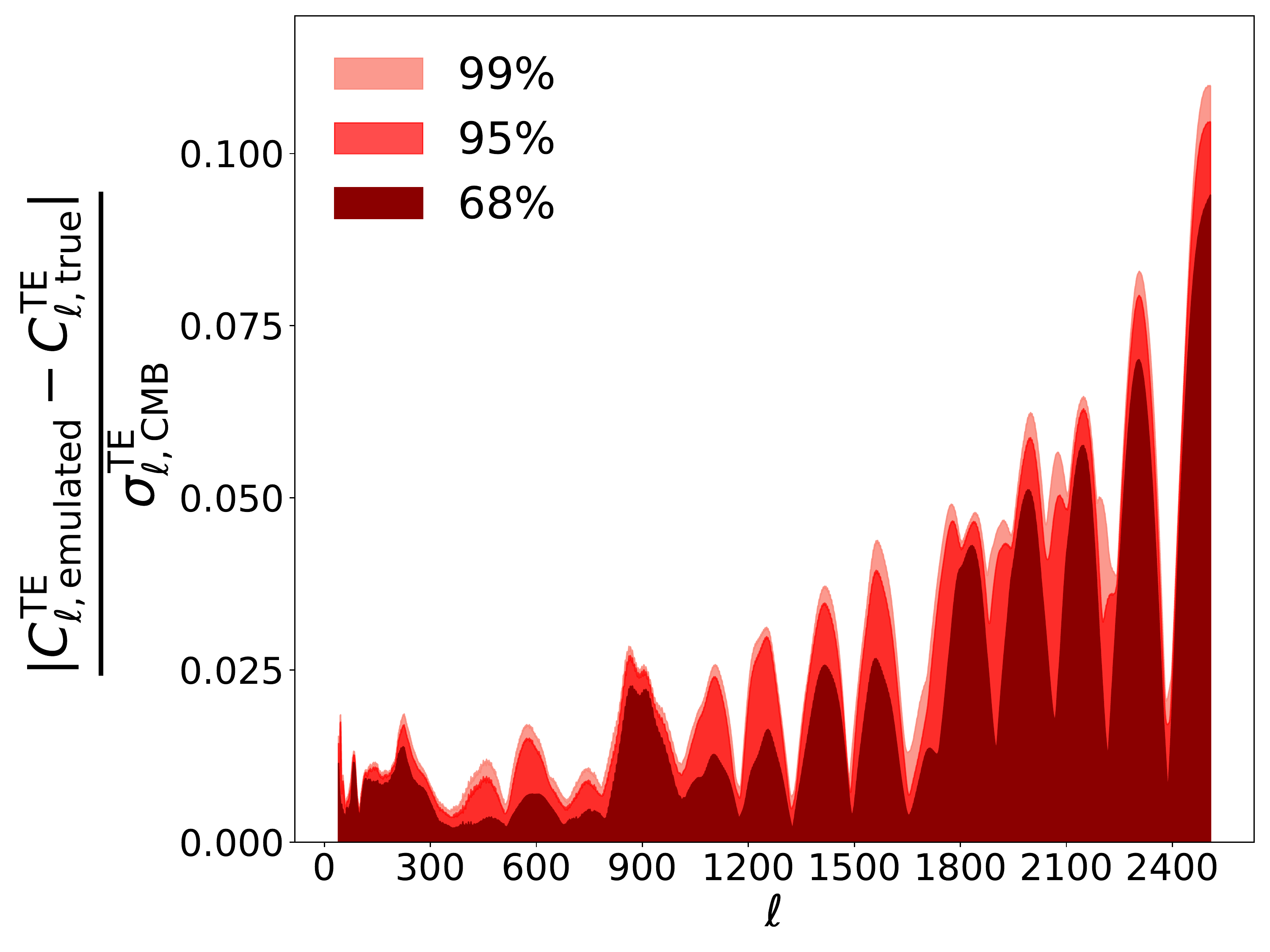}
            \caption[]%
            {{\small $C_{\ell}^{\textrm{TE}}$}}    
            \label{fig:planck_te_emu_narrow}
        \end{subfigure}
        \hfill
        \begin{subfigure}[b]{0.475\textwidth}   
            \centering 
            \includegraphics[width=\textwidth]{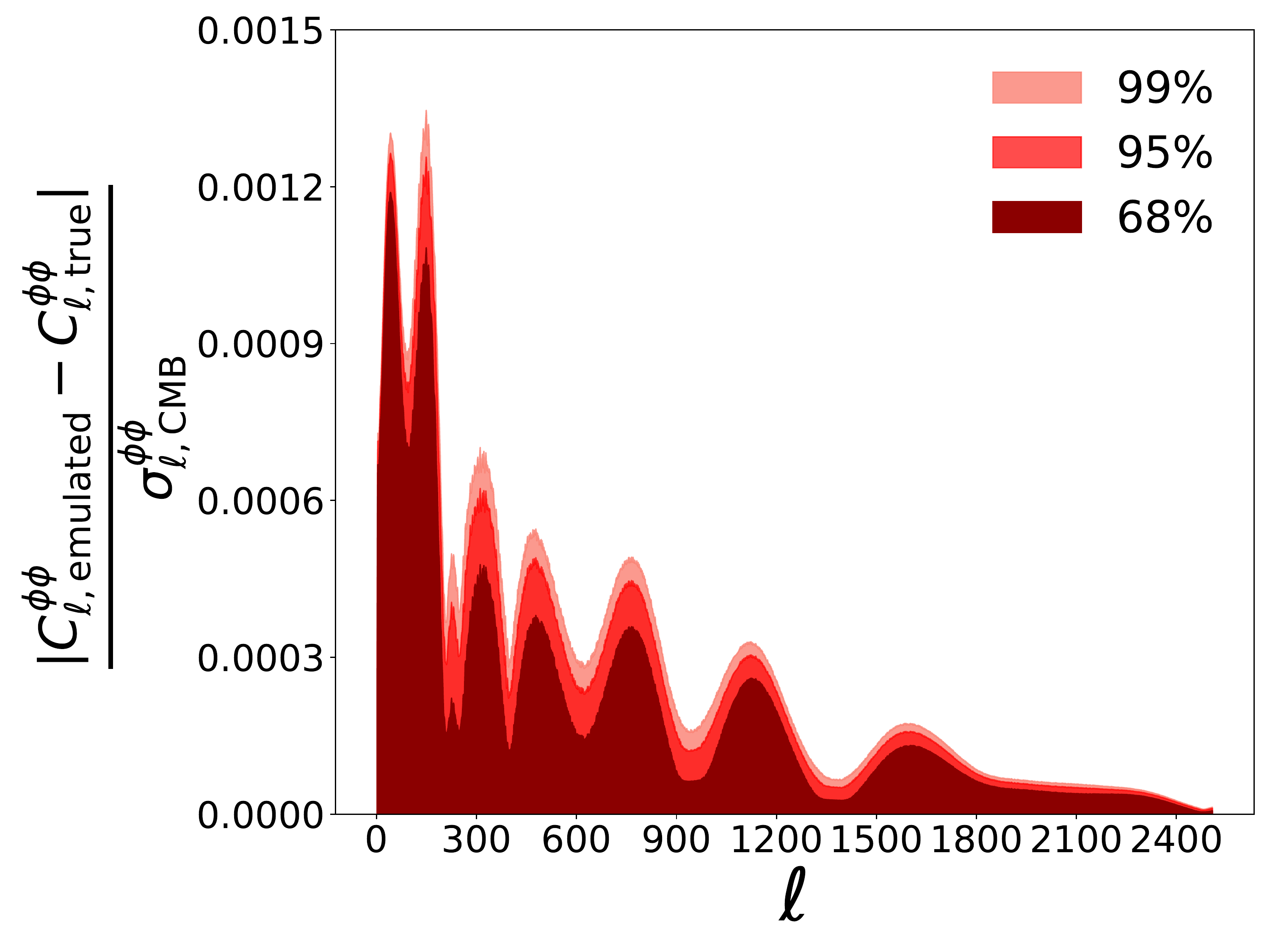}
            \caption[]%
            {{\small $C_{\ell}^{\phi \phi}$}}    
            \label{fig:planck_pp_emu_narrow}
        \end{subfigure}
        \caption[ ]
        {\small  CMB power spectra emulation accuracy on the 5$\sigma$ range test set for a) the temperature power spectrum, b) the polarisation power spectrum, c) the temperature-polarisation cross power spectrum, d) the lensing potential power spectrum. The emulation error is defined with respect to both instrumental and statistical noise, and is defined in Eqs.~(\ref{eq:cl_cmb_error}-\ref{eq:cl_cmb_error_te}). \textit{Dark red}, \textit{red} and \textit{salmon} areas enclose the 68, 95 and 99 percentiles of the test set. Details of the neural models are reported in Appendix~\ref{app:nn}.}. 
        \label{fig:planck_emu_narrow}
    \end{figure*}

\begin{table}
  \centering 
  \begin{tabular}{c|c}
    \textbf{Parameter}               &  \textbf{Prior range}   \\
    \hline
    \hline
    $\omega_{\mathrm{b}}$            &   [0.005, 0.04]    \\
    \hline
    $\omega_{\mathrm{cdm}}$          &    [0.001, 0.99]        \\
    \hline
    $h$                              &    [0.2, 1.0]         \\
    \hline
    $\tau_{\mathrm{reio}}$                              &    [0.01, 0.8]         \\
    \hline
    $n_s$                            &        [0.7, 1.3]      \\
    \hline
    $\mathrm{ln}10^{10}A_s$          &      [1.61, 5]       \\
    \hline
    $A_{\mathrm{Planck}}$            &      $\mathcal{N}(1, 0.0025)$       \\
    \hline
    \hline
    \end{tabular}
   \caption{Prior ranges for the \textit{Planck} analysis. Prior distributions are all taken to be uniform across these ranges, except for the nuisance parameter $A_{\mathrm{Planck}}$ which has a Gaussian prior with mean 1 and standard deviation 0.0025. The ranges on the cosmological parameters correspond to the ranges of validity of our emulators (cf. Table \ref{tab:emu_ranges}).}
  \label{tab:prior_planck}
\end{table}

We validate our emulators using two sets containing  $\sim 2 \cdot 10^4$ spectra each. The first one corresponds to cosmological parameters sampled from a restricted range, corresponding to 5$\sigma$ intervals around the best fit values from \textit{Planck} \citep{Planck18}; the results are reported in Fig.~\ref{fig:planck_emu_narrow} for all probes. As one can see, the distribution of the quantiles is very tight, and almost always 99\% of the spectra are within less than 0.1$\sigma_{\textrm{CMB}}$ at all $\ell$ values.

The second set of spectra corresponds to the same range the emulators were trained on, i.e. the intervals in Table~\ref{tab:emu_ranges}; the results are reported in Appendix~\ref{app:cmb_large}. Unsurprisingly, the errors are slightly larger, reaching 0.2$\sigma_{\textrm{CMB}}$; however, in the next section we show that this level of accuracy for spectra emulation is sufficient to provide accurate and unbiased inference of cosmological parameters in a CMB analysis.

\subsection{Validation on the Planck 2018 likelihood}\label{sec:planck}
\begin{figure*}
    \centering
    \includegraphics[scale=0.4]{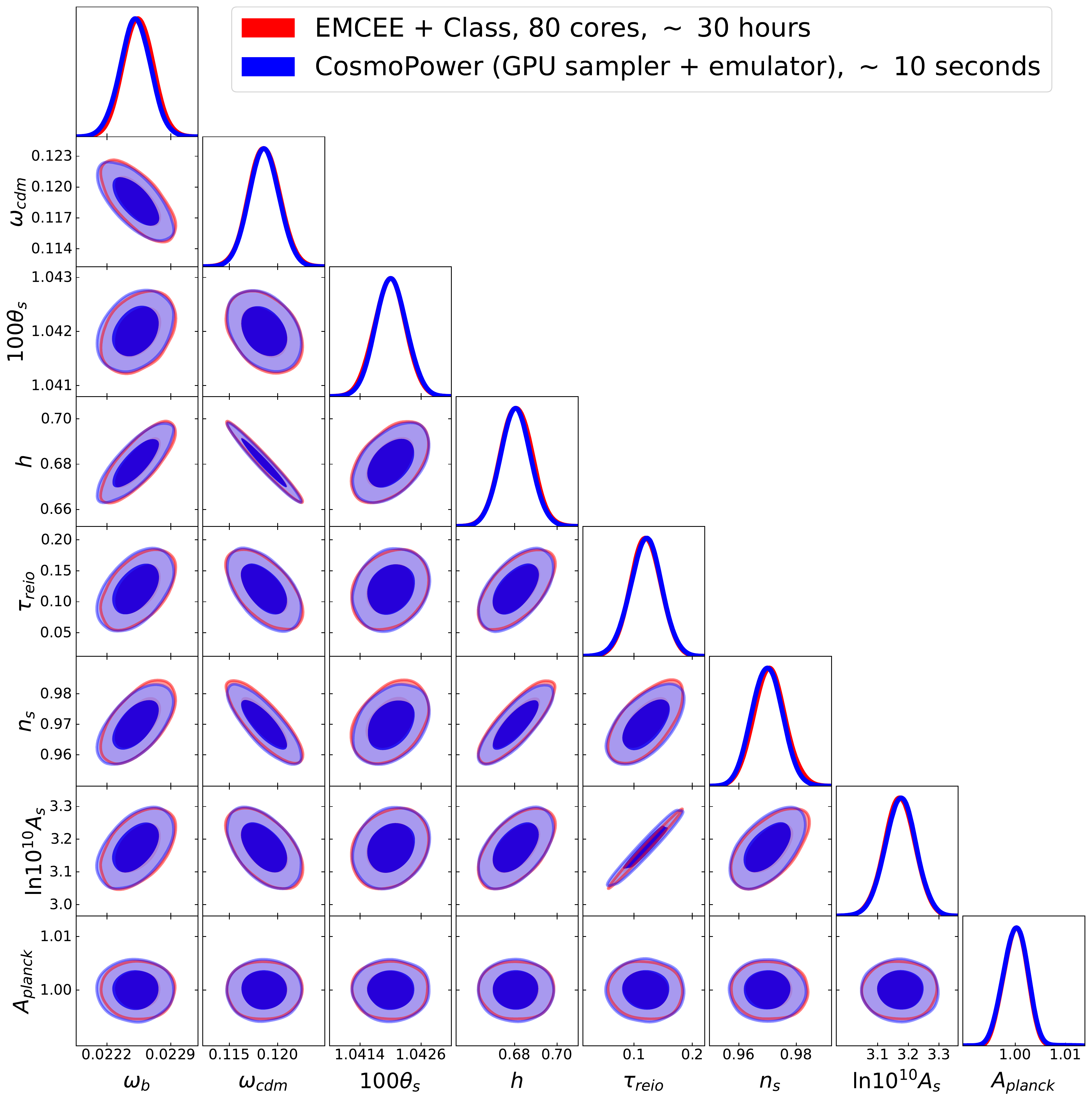}
    \caption{Planck 2018 3x2pt analysis considering $C_{\ell}^{\textrm{TT}}$, $C_{\ell}^{\textrm{EE}}$ and $C_{\ell}^{\textrm{TE}}$. The red contours are obtained in $\sim 1.2 \cdot 10^5$ seconds on 80 CPU cores using \textsc{Class}, while the blue contours take roughly 10 seconds on a single GPU using our neural emulators. Note that the constraints on $100 \theta_{\textrm{S}}$ are derived from the rest of the samples using a Gaussian Process.}
    \label{fig:planck_contours}
\end{figure*}

After assessing the accuracy of our emulators by looking at the residuals of their predictions on the testing set, we performed the final validation check by using the emulators to speed up parameter estimation in a \textit{Planck} CMB inference pipeline \citep{Planck18}. We considered the pure \textsc{Python} implementation of the \textsc{Planck 2018 plik-lite} likelihood available from \citet{Prince19}\footnote{\url{ https://github.com/heatherprince/planck-lite-py}. Note that a similar pure \textsc{Python} implementation is available in the cosmological sampler \textsc{Cobaya}, \url{https://github.com/CobayaSampler/cobaya/blob/master/cobaya/likelihoods/base_classes/planck_pliklite.py}}, which is pre-marginalised over a series of nuisance parameters. The only remaining calibration parameter is a multiplicative correction factor $A_{\mathrm{Planck}}$. The prior ranges for all of the parameters varied in the analysis are reported in Table \ref{tab:prior_planck}.

To further showcase the strength of \textsc{CosmoPower}, to draw from the posterior distribution we use a parallelised affine-invariant MCMC sampler\footnote{\url{https://github.com/justinalsing/affine}}, that is a parallel, GPU-compatible \textsc{TensorFlow} implementation of a variant of the algorithm underlying the \textsc{emcee} sampler  \citep{ForemanMackey13}, based on \citet{Goodman10}. This sampler fully exploits the neural emulators by allowing large batches of likelihood calls to be performed in parallel. Using a single GeForce GTX 1080 GPU, we can obtain the full contours in $\sim 10$ seconds, while the contours using \textsc{Class} and \textsc{emcee} take a total wall-clock time of $\sim 1.2 \cdot 10^5$ seconds using 80 CPU cores, for a final speedup of $\mathcal{O} (10^4)$. Moreover, we train a single GP, as described in Sect.~\ref{sec:methods}, to derive constraints on $100 \theta_{\textrm{S}}$ as well, in the same fashion of the $\sigma_8$ contours obtained in Sect.~\ref{sec:LSS}. We show the posterior contours in Fig.~\ref{fig:planck_contours}.

To obtain log-evidence estimates we re-run both likelihoods with \textsc{MultiNest} and find $\mathrm{log} \mathcal{Z} = -313.72\, \pm 0.15$ and $\mathrm{log} \mathcal{Z} = -313.79 \, \pm 0.15$ for the run with \textsc{Class} and \textsc{CosmoPower}, respectively.


\section{Conclusions}\label{sec:conclusions}

We presented \textsc{CosmoPower}, a suite of cosmological power spectra emulators developed to accelerate by orders of magnitude parameter estimation from Large-Scale Structure (LSS) and Cosmic Microwave Background (CMB) surveys. \textsc{CosmoPower} emulates matter and CMB power spectra computed by Boltzmann codes such as \textsc{Camb} and \textsc{Class}. Sourcing power spectra from Boltzmann codes is the computational bottleneck for two-point statistics analyses of cosmological fields; \textsc{CosmoPower} bypasses this step, providing orders-of-magnitude acceleration to the inference pipeline. Power spectra emulation is performed using a neural network (NN) to parameterise the mapping between cosmological parameters and power spectra or their Principal Component Analysis (PCA) coefficients. 

In this paper, we presented emulators for the linear and nonlinear matter power spectrum, as well as for the CMB temperature, polarisation and lensing power spectrum. We showcased the performance of \textsc{CosmoPower} with applications to Stage III and simulated Stage IV surveys, including: a 3x2pt and cosmic shear analysis of the Kilo-Degree Survey (KiDS), a mock \textit{Euclid}-like cosmic shear analysis and a \textit{Planck} 2018 joint temperature and polarisation analysis. In all of these cases, the power spectra emulators provided unbiased cosmological inference, at a fraction of the time required by the same pipelines run with power spectra sourced from Boltzmann solvers. In the following, we summarise the main properties of \textsc{CosmoPower}, compare its performance with that of other emulators, and discuss some of its planned future extensions.

\subsection{Key properties of \textsc{CosmoPower}}

The following key properties of \textsc{CosmoPower} make it an invaluable tool for application to future Stage IV analyses.

\begin{itemize}

\item \textbf{Speed-up}. First and foremost, the use of \textsc{CosmoPower} to replace Boltzmann codes in likelihood evaluations provides an impressive speed-up factor. In the applications considered in this paper, \textsc{CosmoPower} provided an acceleration factor up to $O(10^4)$ with respect to standard analyses with Boltzmann codes. These numbers refer to the full inference pipeline; if we restrict to timing a single power spectrum evaluation, the speedup increases even further up to $O(10^5)$, respectively. These numbers are expected to increase as we extend \textsc{CosmoPower} to cosmologies beyond the flat $\Lambda$CDM model which was assumed throughout this paper. In this sense, the acceleration quoted in this analysis is to be regarded as a lower bound for the speed-up achievable with \textsc{CosmoPower}.

\smallskip

\item \textbf{GPU acceleration}. Our emulators are based on neural networks implemented in \textsc{TensorFlow}. As such, they benefit from an additional speed-up when run on Graphics Processing Units (GPU) or Tensor Processing Units (TPU)\footnote{both freely available on Google Colab \url{http://colab.research.google.com/}. In the \textsc{CosmoPower} GitHub repository we provide \textsc{Jupyter} notebook examples to run \textsc{CosmoPower} on Colab GPU.}.

The speed-up calculated for the full inference pipeline differs from the one computed for a single power spectrum evaluation because a single likelihood evaluation is slower than a power spectrum evaluation: computing the projected angular spectra of the LSS probes, or binning the CMB spectra as performed in the \textit{Planck} likelihood, requires a series of numerical operations. These parts of the likelihood evaluation are computationally intensive regardless of the method employed to source power spectra. This is particularly true for Stage IV LSS surveys, which will have a high number of redshift bins and hence will require the computation of a high number of bin cross-correlations (cf. Eq.~\ref{eq:limber}). These expensive loops in the likelihood evaluation can be massively accelerated by implementing the likelihood in \textsc{TensorFlow} or \textsc{JAX} \citep{Jax18} and running the inference pipeline on GPUs.

In this paper we showed how running \textsc{CosmoPower} in a pure \textsc{TensorFlow}-based version of the \textit{Planck} likelihood, embedded within a framework for cosmological inference also implemented in \textsc{TensorFlow}, provided contours in $\sim$ 10 seconds. Running an inference pipeline with the \textit{Planck} likelihood is a notoriously computationally intensive task: this example of speed-up achieved with a \textsc{Tensorflow}-based likelihood and power spectra clearly shows how beneficial the combination of highly optimised software for power spectra emulation and Bayesian posterior sampling can be. We advocate for moving towards implementations of cosmological inference software in \textsc{TensorFlow} or \textsc{JAX}, to fully exploit the power of highly optimised software that can be run on GPU and is auto-differentiable. The reason for this is the high number of nuisance parameters that is expected to be required to model the observables, in addition to the large size of the data vector. Particularly, if one desires to make use of the unprecedented amount of data provided by these surveys to investigate beyond-$\Lambda$CDM cosmologies, the implementation of cosmological inference frameworks leveraging GPU acceleration is of the utmost importance. \textsc{CosmoPower} aims at providing such a framework, incorporating not only trained emulators but also template \textsc{TensorFlow}-based likelihoods for LSS and CMB surveys that can be easily adapted for application to different surveys.

\smallskip

\item \textbf{Full differentiability}. \textsc{CosmoPower} provides emulators for the power spectra that are based on neural networks and implemented in \textsc{TensorFlow} \citep{Abadi15}. Thus, these emulators are by construction fully differentiable, a feature which makes them ideal for gradient-based inference, such as Hamiltonian Monte Carlo \citep{Hajian07}. If desired, they can also be used for instantaneous Fisher matrix computation and linear data compression with e.g. the MOPED algorithm \citep{Heavens00}, leveraging the possibility of calculating derivatives with respect to the input parameters by auto-differentiation.

\smallskip

\item \textbf{Accuracy}. The procedure followed to validate the accuracy of our emulators guarantees that they can be safely used for analyses of Stage IV surveys. Crucially, we verified this statement not only by checking the residuals between emulated and real power spectra in the testing set, but also by validating our emulators with full posterior inference analyses from state-of-the-art surveys, as well as from simulated Stage IV surveys. In carrying out this comparison at the contours level, we performed an additional validity check between \textsc{CosmoPower} and the Boltzmann codes \textsc{Camb} and \textsc{Class}. To train our models we used power spectra generated with \textsc{Camb}; instead, when comparing the \textsc{CosmoPower} contours against those obtained from a traditional inference pipeline sourcing power spectra from a Boltzmann code, we used \textsc{Class} for the latter. As shown in Figs.~\ref{fig:kidsxgama_contours}, \ref{fig:euclid_contours} and \ref{fig:planck_contours}, contours obtained with the emulators (trained on \textsc{Camb}) were always found in excellent agreement with the contours obtained from the inference pipelines sourcing power spectra from \textsc{Class}, including in the simulated Stage IV \textit{Euclid}-like configuration. We also verified that replacing \textsc{Class} with \textsc{Camb} in the inference pipelines provides contours with a similar level of agreement: this means that the difference between \textsc{CosmoPower} predictions and Boltzmann-computed power spectra are not bigger than the differences between power spectra computed with different Boltzmann codes.

\smallskip

\item \textbf{Parameter range}. The parameter range over which our models are trained is very large, covering the full Planck prior range in the CMB case and the full KiDS-1000 prior range in the LSS case. The combination of high accuracy and wide validity range allows the user of \textsc{CosmoPower} to safely replace Boltzmann codes with our emulators when computing power spectra, even for those practical applications where high accuracy over broad prior ranges is crucial, such as posterior predictive cross-validation. The accuracy of our emulators even in extreme regions of the parameter space considered for their training is confirmed by the good agreement between log-evidence values obtained in the likelihood runs with power spectra sourced from \textsc{Class} and \textsc{CosmoPower}.

\smallskip

\item \textbf{Flexibility}. By construction, \textsc{CosmoPower} emulates cosmological power spectra taking in input only those cosmological parameters that are part of the mapping between input cosmologies and output power spectra. This means that, for example, in the LSS case, the key emulated quantity is the matter power spectrum and not the cosmic shear, galaxy-galaxy lensing or galaxy clustering projected power spectra. The rationale behind this choice is that the angular spectra of cosmological probes are quantities derived from the matter power spectrum, by integrating it over a kernel which depends on the redshift distributions. In addition, contaminant terms due to e.g. intrinsic alignments are also obtained by integration of the matter power spectrum, and modulated by nuisance parameters which are not part of the cosmological model. By avoiding to include those additional parameters in the target mapping for the emulator, \textsc{CosmoPower} acquires a unique flexibility that makes it completely independent of astrophysical nuisance parameters, such as intrinsic alignment and galaxy bias parameters, that do not modify the matter power spectrum prediction. This means that our emulators do not need to be trained for different choices of these astrophysical parameters. In particular, no re-training is required if one wishes to implement different prescriptions for the modelling of contaminants such as intrinsic alignments, e.g. by inserting additional nuisance parameters, as long as those parameters do not modify the prediction for the matter power spectrum. A similar argument is applicable to the modelling of redshift distributions, which will likely require even more nuisance parameters than the mean shifts used in our simulated \textit{Euclid}-like analysis \citep[see e.g.][]{Hasan21}. As an additional bonus, emulating the 3D matter power spectrum will allow us to investigate in future work the use of \textsc{CosmoPower} for emulating cosmological power spectra beyond the Limber approximation (see below).

\smallskip

\item \textbf{Linear and nonlinear power spectra}. \textsc{CosmoPower} provides emulators for both linear power spectra and nonlinear correction factor. For the latter, the \textsc{HMcode} prescription is currently implemented in the emulator. The \textsc{HaloFit} model (which \textsc{HMcode} is based on) is also available to the user. The separation between linear power spectrum and nonlinear correction factor is particularly useful as it allows us to integrate in \textsc{CosmoPower} additional models for nonlinearities as they become available. On the linear level, modified Boltzmann codes for beyond-$\Lambda$CDM models exist \citep[for example \textsc{HiClass} for Horndeski models;][]{Zumalacarregui17} that provide linear predictions for the matter power spectrum in these extended cosmologies. As we extend \textsc{CosmoPower} to these models, we can add new emulators for linear power spectra trained on these modified Boltzmann codes.   

\smallskip

\item \textbf{``Train-once-use-repeatedly'' approach and interface with cosmological samplers}. While we provide all the tools necessary to repeat the training if desired, we stress that this operation has already been performed and does not need to be repeated, as long as the emulators are employed assuming the cosmological model and range indicated in Table~\ref{tab:emu_ranges}. In addition, \textsc{CosmoPower} can be called from all commonly used cosmological samplers. In this paper, for example, we used \textsc{CosmoPower} within the cosmological samplers \textsc{MontePython} and \textsc{Cobaya}. The user of \textsc{CosmoPower} simply needs to write a likelihood for the LSS or CMB survey considered, and replace the call to the Boltzmann code, necessary to obtain the matter or CMB power spectra, with a call to \textsc{CosmoPower}.

\smallskip

\item \textbf{Derived parameters}. Emulators developed in the literature usually provide a fixed parameterisation to emulate from. For example, if an emulator is trained using $\mathrm{ln} 10^{10} A_s$ it is not possible to get a prediction for a corresponding value of $\sigma_8$ or $S_8$. \textsc{CosmoPower} provides emulators trained on different combinations of parameters. For example, the 3x2pt KiDS-450+GAMA analysis was performed with an emulator trained on $\mathrm{ln} 10^{10} A_{\textrm{s}}$ as input parameter, while the KiDS-1000 analysis used $\sigma_8$ in input. In addition, \textsc{CosmoPower} also allows the user to post-process a sampled chain to obtain very efficiently derived parameters that were not originally sampled. Moreover, we provide Gaussian Processes (GPs) to obtain derived parameters that were not used as input to the emulators. The accuracy of these GPs was tested not only on a test set, but also against the actual contours obtained with the Boltzmann codes.

\end{itemize}

\subsection{Comparison with previous work}
Here we compare our emulators to other existing approaches to power spectra emulation. We start by noticing that \textsc{CosmoPower} provides an emulation framework for \textit{both} LSS and CMB. To our knowledge, this is a unique feature, only partially shared by \textsc{PICO} and \textsc{CosmoNet} (both, however, emulating matter transfer functions rather than power spectra). These two packages are not actively maintained nor trained with the same accuracy or across the same parameter ranges, which limits their applicability to Stage IV analyses. As far as the matter power spectrum is concerned, the methods closest to ours in terms of emulation are the one implemented in \citet{Arico21}, even though limited to the linear power spectrum, and \citet{Agarwal12}, limited to \textsc{HaloFit} nonlinearities.  

We note that applying the emulator to a complete inference analysis from a simulated Stage IV survey, as done in our paper, is a necessary step to ensure that the newly developed tool can be safely applied in practical analyses. On the contrary, checking residuals in the testing set between predicted and real spectra is not a sufficient accuracy test. While an emulator may be performing with e.g. sub-percent accuracy at the level of residuals, this may still not be enough to retrieve unbiased cosmological contours, as we verified first-hand while testing \textsc{CosmoPower}. This is due to the fact that the accuracy threshold for the emulation can only be defined by the specific application for which these emulators are designed. In other words, it is the inference pipeline that dictates the accuracy threshold to be met by the emulator. In general, parameter estimation in Bayesian inference pipelines requires a certain level of accuracy in the observables computed, which in the specific case of cosmological two-point statistics analyses reflects into certain accuracy requirements in the power spectra computed by Boltzmann codes. Hence, we argue that the principled approach to validate an emulator accuracy is to compare its performance within an inference pipeline for a target experiment, which in our case is a Stage IV survey configuration. Note that, while testing \textsc{CosmoPower}, we experienced first-hand that emulators performing greatly on Stage III experiments failed in producing equally correct contours on a simulated Stage IV survey.

GP-based emulators such as the ones used in \citet{Mootoovaloo20, Mootoovaloo21, Ramachandra20, Ho21} require fewer training samples than a NN emulation framework like \textsc{CosmoPower}; however, GPs also provide reduced speed-ups compared to NNs. On the other hand, GPs also provide a way to propagate the uncertainty in their prediction to the final posterior distribution, whereas simple NNs like those implemented in this version of \textsc{CosmoPower} lack this feature. In future versions of \textsc{CosmoPower} we will investigate the use of Bayesian Neural Networks or ensemble NN predictions for this purpose.

\citet{Mootoovaloo20} developed GP emulators of cosmic shear band powers for the KiDS-450 survey, with the option of compressing the band powers into MOPED coefficients and learning those coefficients with the GP instead of the band powers themselves. Similarly, \citet{ManriqueYus19} developed NN emulators of the 3x2pt angular power spectra. Both of these approaches are constrained by the choice of the redshift distributions specific to the survey, which enters the expression of the angular power spectra. In addition, the method of \citet{Mootoovaloo20} also relies heavily on the choice of nuisance parameters used to model the power spectra; these parameters need to be `learnt' by their GP. \textsc{CosmoPower} is free from any of these restrictions: targeting emulation of the matter power spectra, our emulators are completely flexible to be used for any redshift distribution and choice of nuisance parameters. While the speed-up obtained by emulating the matter power spectrum may be smaller than that obtained from emulating the angular power spectra of the different probes, we believe this computational overhead can be avoided by rewriting the likelihood of interest in \textsc{TensorFlow} and running it on a GPU together with the emulators.

Finally, \citet{Albers19} implemented an interesting NN-based acceleration of source functions in the calculation of CMB power spectra within the Boltzmann code \textsc{Class}. Emulating source functions provides great flexibility in the pipeline, as for example it allows one to compute higher-order correlators and non-linear transfer functions without retraining. On the other hand, emulation of full power spectra performed in \textsc{CosmoPower} provides greater speed-ups and is better suited for implementation of full inference pipelines on GPUs.

\subsection{Future work}
\textsc{CosmoPower} is an open-source package provided to the cosmological community as a tool to accelerate intensive computations within Bayesian inference pipelines of LSS and CMB surveys. In this paper we considered the emulation of power spectra, which represents the bottleneck for two-point statistics analyses of cosmological fields. However, this paper also marks the starting point of a longer-term project, with the goal of extending the \textsc{CosmoPower} framework to accelerate the forward-modelling of multiple cosmological observables with machine learning.

\begin{itemize}
    \item \textbf{Higher-order statistics, systematics and beyond-Limber spectra}. We plan to train emulators for higher-order statistics such as the bispectrum. Not only does the bispectrum contain complementary cosmological information to the power spectrum \citep[see e.g.][]{Pyne21}, but it is also required to calculate (computationally expensive) corrections to the power spectrum as those arising from dropping the reduced shear approximation \citep{Deshpande20}. In addition, multiple observational systematics in cosmic shear analyses \citep[see e.g.][]{Paykari20} can be modelled with machine learning techniques and their effect on cosmological parameter estimation can thus be properly accounted for. Finally, for LSS a key advantage within our emulator is given by targeting the matter power spectrum, as opposed to the angular power spectra of the cosmological probes. This choice will allow us to investigate efficient computation of non-Limber projected quantities in future work.

\smallskip

   \item \textbf{Beyond-$\Lambda$CDM cosmologies}. As already mentioned above, we plan to extend \textsc{CosmoPower} to models beyond the flat $\Lambda$CDM one considered in this paper. For example, emulation of power spectra in non-flat cosmologies is of the utmost importance, since their calculation by Boltzmann codes are computationally intensive \citep[see e.g.][]{Handley21}.
   More generally, power spectra computed in alternative cosmologies are considerably more demanding to compute for Boltzmann codes than in $\Lambda$CDM. Instead, we expect NN architectures similar to the ones considered in this paper to be equally accurate for emulating beyond-$\Lambda$CDM spectra (while possibly requiring larger training sets). Consequently, the evaluation time of these beyond-$\Lambda$CDM emulators will remain essentially the same as those reported in this paper. This will produce an even greater speed-up factor over Boltzmann codes. Some of these extensions to $\Lambda$CDM are already being developed and will be presented in a future publication (Spurio Mancini et al., in prep.).

\smallskip

    \item \textbf{A fully differentiable cosmology library}. In the longer term, \textsc{CosmoPower} will be extended to provide a completely differentiable library for cosmological computations. This will also include much simpler functions to emulate, such as cosmological distances. As we move towards the era of Stage IV surveys, the statistical challenges in analysing those datasets will certainly require increased sophistications in the Bayesian inference engines available. Differentiability is key in unlocking the possibility of efficient gradient-based inference, a promising avenue to tackle the challenge represented by the high-dimensional parameter spaces characterising the analyses of Stage IV surveys. Therefore, endowing the cosmological community with a fully differentiable forward model of multiple observables is a task of paramount importance, which we aim to accomplish with \textsc{CosmoPower}.

\smallskip

   \item \textbf{Interpretable machine learning}. Alternative methods for compression and emulation of cosmological quantities, such as autoencoders and symbolic regression \citep{Udrescu20}, will be explored within \textsc{CosmoPower}, with the goal of maximising the computational efficiency and interpretability of the machine learning framework developed.
\end{itemize}

\textsc{CosmoPower} is a tool that allows for principled, non-invasive application of machine learning within a rigorous Bayesian framework for uncertainty quantification. We are confident that emulation techniques like those presented here will greatly enhance the scientific return from Bayesian inference analyses of upcoming Stage IV surveys.

\section*{Data Availability}
\textsc{CosmoPower} is freely available at \url{https://github.com/alessiospuriomancini/cosmopower}. 

\section*{Acknowledgements}
We thank S. Joudaki for a thorough review of the manuscript and H. Peiris, A. Pourtsidou for useful discussions. We thank the anonymous referee for a meticulous and insightful review of this paper. ASM is supported by the MSSL STFC Consolidated Grant. DP is supported by the STFC UCL Centre for Doctoral Training in Data Intensive Science. JA is supported by the research project grant \emph{Fundamental Physics from Cosmological Surveys} funded by the Swedish Research Council (VR) under Dnr 2017-04212. Generation of the data used in this work has been performed in part on the Beaker and Hypatia clusters at UCL. We thank Edd Edmondson for technical support. This work has been partially enabled by funding from the UCL Cosmoparticle Initiative. We acknowledge the use of \textsc{GetDist} \citep{Lewis19} to obtain corner plots.

Based on data products from observations made with ESO Telescopes at the La Silla Paranal Observatory under programme IDs 177.A-3016, 177.A-3017 and 177.A-3018. Based on observations obtained with Planck (\url{http://www.esa.int/Planck}), an ESA science mission with instruments and contributions directly funded by ESA Member States, NASA, and Canada.

\bibliographystyle{mnras}
\bibliography{references} 

\appendix

\section{Emulation details}\label{app:methods}
\subsection{Neural network}
\label{app:nn}
A neural network (NN) consists of a sequence of layers connecting some input to some output, in this case cosmological parameters to power spectra (or their Principal Components). As depicted in Fig.~\ref{fig:emulation}, each layer performs a linear combination of the previous layer, and applies a nonlinear function to increase the expressiveness of the model. Following \citet{Alsing20}, we choose the following activation function for all hidden layers in our neural networks
\begin{align}\label{eq:act_func}
    f(\boldsymbol{x}) = \left ( \boldsymbol{\gamma} + \left ( 1 + e^{-\boldsymbol{\beta}\odot\boldsymbol{x}} \right ) ^{-1} \odot (1-\boldsymbol{\gamma}) \right ) \odot \boldsymbol{x} \ ,
\end{align}
where $\boldsymbol{\beta}$ and $\boldsymbol{\gamma}$ are optimised together with the rest of the network parameters, and $\odot$ indicates the element-wise product. The activation function in Eq. \ref{eq:act_func} can be seen as a stack of scalar activation functions for each node $\mathrm{j}$ with independent hyperparameters $\beta_{\mathrm{j}}$ and $\gamma_{\mathrm{j}}$. We experimented with other, more traditional activation functions such as the hyperbolic tangent (tanh) or Rectified Linear Unit (ReLU) \citep{Agarap18} and we always found them to perform slightly worse than our custom function in terms of accuracy. 

In the CMB case the output layer is made of 2507 nodes, i.e. one for each multipole $\ell$ from $\ell_{\mathrm{min}}=2$ to $\ell_{\mathrm{max}}=2508$ included. When building the training and testing set, we make sure to ask the Boltzmann code for an explicit calculation of the CMB power spectrum at each one of the multipoles in the range, as this reduces the amount of interpolation performed internally by the Boltzmann code and, in turn, increases the accuracy of the \textsc{CosmoPower} emulation. For the matter power spectrum the output layer is made of 420 nodes, corresponding to the sampled $k$-modes: these are chosen such that the baryon acoustic oscillations features are more densely sampled, in an analogous fashion to what is performed internally by \textsc{class} and \textsc{camb}, albeit in this case with a cosmology-independent sampling.

Our neural networks always use 4 hidden layers of 512 neurons each, for both CMB and LSS; each neuron is associated to a weight and a bias, as described in Sect.~\ref{sec:methods} and Fig.~\ref{fig:emulation}. We did not perform a fully exhaustive optimization of hyperparameters of the neural networks such as the number of hidden layers or neurons. In our experiments we typically found that an architecture with at least three hidden layers of a few hundred neurons each was needed to achieve high accuracy results, with a four-layer configuration being even more performing. It is possible that equally, if not more accurate results may be achieved with a smaller architecture, albeit with more training samples. However, handling such large datasets may be a computationally non-trivial challenge, particularly in terms of memory requirements.

The network's parameters are optimised using Adam \citep{Kingma14} with default parameters, and the loss function to minimise is chosen to be the mean squared error between the emulated and the true power spectra. We keep apart 20\% of the training set for validation purposes. The learning rate is initially set to $1\cdot10^{-2}$, and then decreased by a factor of 10 each time the validation loss does not decrease for 20 epochs, where each epoch corresponds to feeding the whole training set into the network; the final learning rate is $1\cdot10^{-6}$. The batch size is changed accordingly, starting from $1\cdot10^3$, then $1\cdot10^4$, up to $5\cdot10^4$.

\subsection{Principal Component Analysis}
\label{app:pca}
We compared the performance of the direct NN mapping with an alternative emulation method where the spectra are first compressed to their Principal Components, following \textsc{Speculator} \citep{Alsing20}, which we refer to for the full details. Note that this is necessary for the $C_{\ell}^{\textrm{TE}}$ case, as due to the dynamic range of these spectra it is not possible to consider its logarithmic features; moreover, we found the performance of this second emulator superior over the direct NN mapping when emulating $C_{\ell}^{\phi \phi}$.

We keep 512 and 64 Principal Components for $C_{\ell}^{\textrm{TE}}$ and $C_{\ell}^{\phi \phi}$, respectively; the neural network is then trained in the same way as described in Appendix \ref{app:nn}. To obtain a power spectrum, we map the cosmological parameters to the predicted Principal Components, and then use the learnt change of base to map the Principal Components into the predicted power spectrum.

\subsection{Gaussian Process}
\label{app:gp}
We use a Gaussian Process (GP) to obtain derived cosmological parameter constraints from given samples of the posterior distribution of the other parameters. When obtaining a derived parameter $\theta_{\textrm{der}}$, we assume that there exists a mapping $\theta_{\textrm{der}} = f(\boldsymbol{\theta})$, where $\boldsymbol{\theta}$ indicates the parameters the emulator was trained on, and model $f(\boldsymbol{\theta})$ as a GP.  Following \citet{SpurioMancini20}, we assume that the function $f(\boldsymbol{\theta})$ follows a normal distribution with zero mean and a parametric covariance matrix, usually referred to as \textit{kernel} $K \left( \boldsymbol{\theta},\boldsymbol{\theta'} ; \boldsymbol{\psi} \right)$, with $\boldsymbol{\theta}$ and $\boldsymbol{\theta'}$ indicating two points in parameter space, and $\boldsymbol{\psi}$ representing the trainable hyperparameters.

We choose the Automatic Relevance Discovery (ARD) $3/2$ Mat$\acute{\mathrm{e}}$rn kernel \citep{Neal96, Rasmussen05}, which is expressed as
\begin{align}
    K_{\mathrm{ARD-Mat}\acute{\mathrm{e}}\mathrm{rn-3/2}} \left( \boldsymbol{\theta}, \boldsymbol{\theta '}; \boldsymbol{\psi} \right) = \alpha^2 \left( 1 + \sqrt{3} r \right) \exp \left(  -\sqrt{3} r \right) \ , 
\end{align}
where
\begin{align}
    r = \sqrt{ \sum_{j=1}^n \frac{ (\theta_j - \theta_j ')^2 }{\sigma_j^2} } \ ,
\end{align}
$n$ is the length of the vector $\boldsymbol{\theta}$ (in this case the number of cosmological parameters), and the hyperparameters $\boldsymbol{\psi} = \{ \alpha, \boldsymbol{\sigma} \} $ are the signal standard deviation $\alpha$ and a characteristic scale $\sigma_j$ for each input parameter $j = 1,  \dots,  n$. We use the software \textsc{GPy}\footnote{\url{http://github.com/SheffieldML/GPy}} to train the GP, i.e. to optimise the hyperparameters of the kernel, using the tuples $\left ( \boldsymbol{\theta}, \theta_{\textrm{der}} \right )$ from the Boltzmann solver \textsc{Camb} as training data.

\section{Validation on the KiDS-1000 likelihood}
\label{app:kids1000}
\begin{figure*} 
  \begin{subfigure}{\columnwidth}
    \centering
    \includegraphics[scale=0.55]{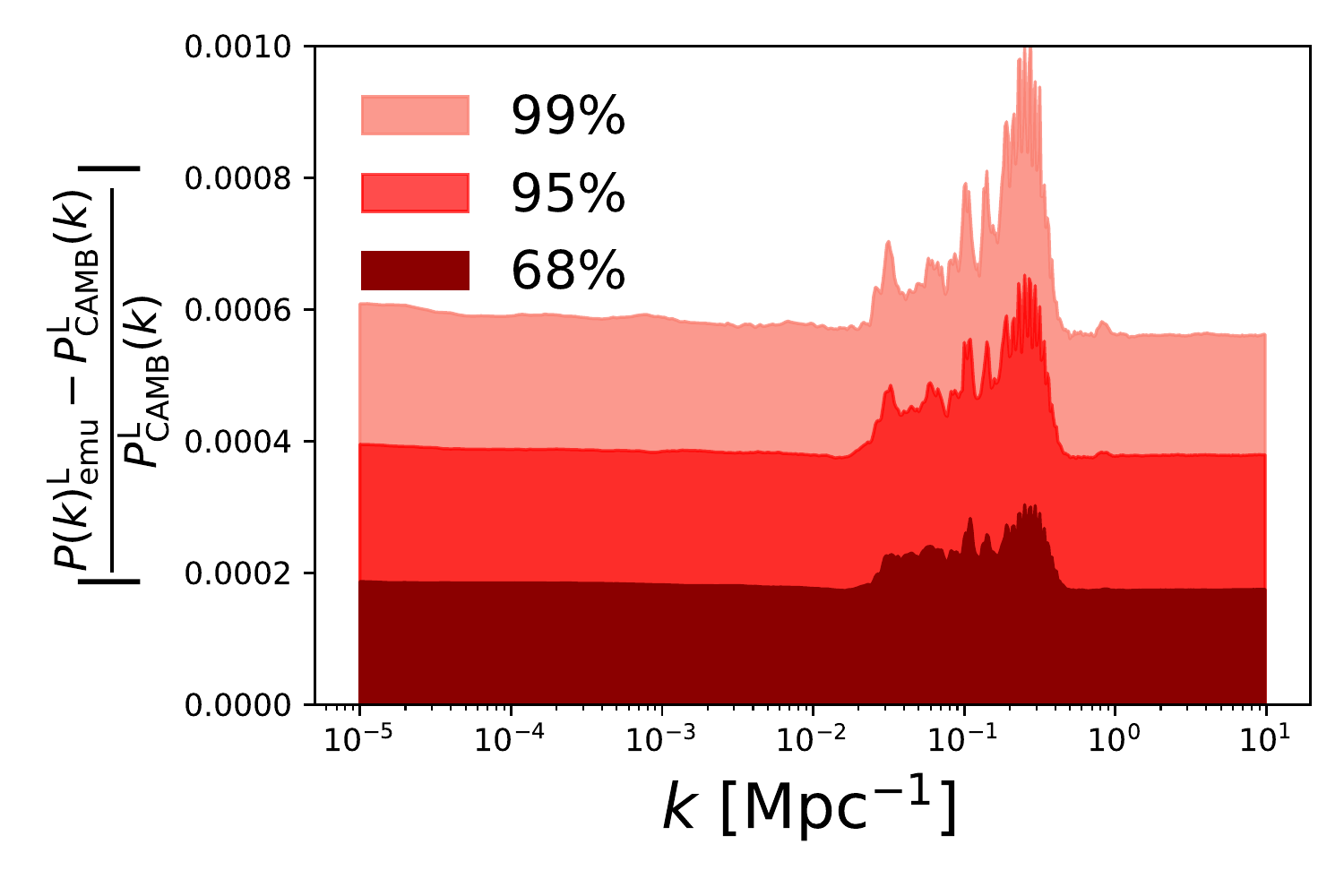}
    \end{subfigure}
  \begin{subfigure}{\columnwidth}
    \centering
    \includegraphics[scale=0.55]{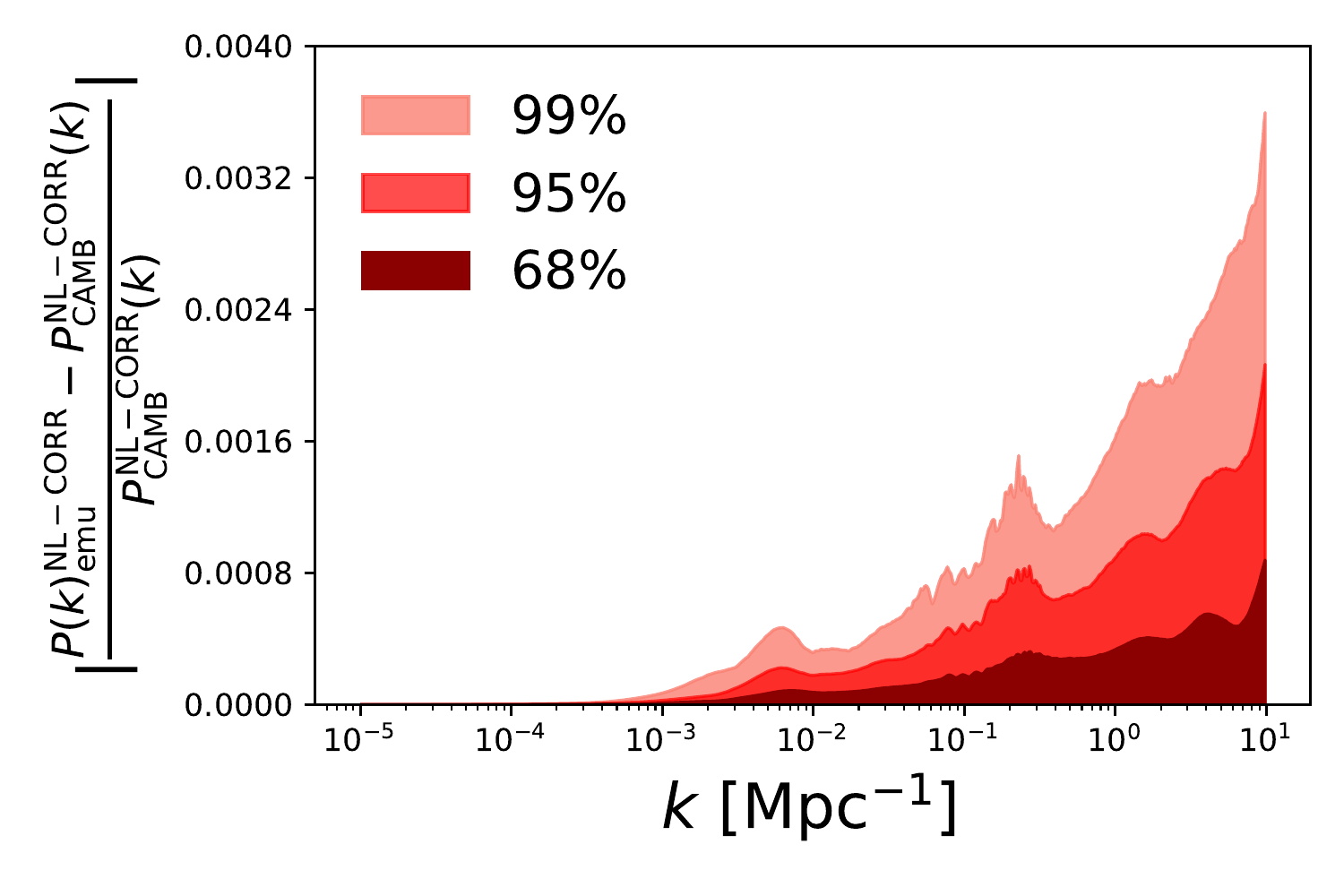}
    \end{subfigure}
  \caption{Same as Fig.~\ref{fig:acc_mnu0} but for a cosmological model with one massive neutrino with mass $m_{\nu} = 0.06$ eV/$c^2$.}
  \label{fig:acc_mnu006}
\end{figure*}

We include an application of \textsc{CosmoPower} to a recent cosmic shear band power analysis of the KiDS survey, covering $\sim$1000 deg$^2$ \citep[KiDS-1000,][]{Asgari20}. For this analysis we need to train a matter power spectrum emulator for a model with one massive neutrino with mass $m_{\nu}=0.06$ eV/$c^2$. We report in Fig.~\ref{fig:acc_mnu006} the same accuracy plots shown in Fig.~\ref{fig:acc_mnu0} for a massless neutrino model. Similar levels of accuracy of the linear power spectrum and nonlinear correction are achieved. In future extensions of \textsc{CosmoPower} we will integrate emulation over the neutrino mass as an additional parameter. Note also that in this analysis, following \citet{Asgari20}, we sample in $S_8$ rather than in $\mathrm{ln} 10^{10}A_{\mathrm{s}}$. 

\begin{figure*}
    \centering
    \includegraphics[scale=0.25]{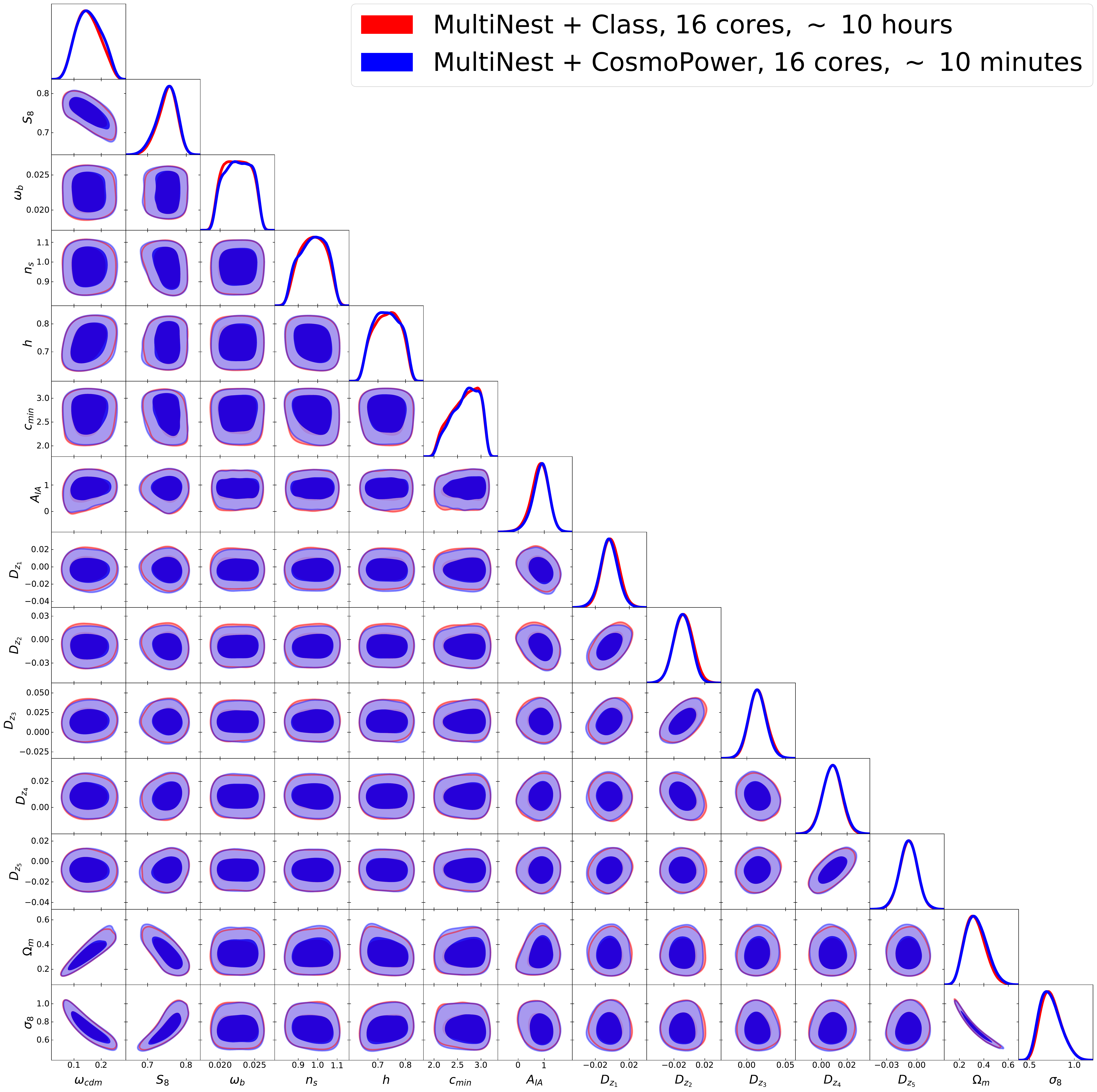}
    \caption{Contours for the KiDS-1000 cosmic shear band power analysis. In \textit{red} contours obtained with \textsc{Class}, in \textit{blue} those obtained with \textsc{CosmoPower}.}
    \label{fig:k1k_contours}
\end{figure*}

The uniform prior ranges assumed for this analysis are the same used in \citet{Asgari20} and they are reported in Table \ref{tab:prior_kids1000}. Following \citet{Asgari20} we include one $z$-shift for each of the five redshift bins, $D_{z_i}$, for $i = 1, \dots, 5$; for these shift parameters we set a correlated Gaussian prior. We refer to \citep{Asgari20, Joachimi20} for further details, in particular on the redshift distributions and covariance matrix modelling. We run the inference pipeline\footnote{adapted from \url{https://github.com/BStoelzner/KiDS-1000_MontePython_likelihood}} within the cosmological sampler \textsc{MontePython}. In Fig.~\ref{fig:k1k_contours} we show the excellent agreement between contours obtained with \textsc{Class} and \textsc{CosmoPower}, with the latter providing a speed-up factor of approximately 50. Values of the log-evidence are also in good agreement: $\mathrm{log} \, \mathcal{Z} = -81.45 \pm 0.14$ and $\mathrm{log} \, \mathcal{Z} = -81.40 \pm 0.14$ for the run with \textsc{Class} and \textsc{CosmoPower}, respectively.
\begin{table}
  \centering 
  \begin{tabular}{c|c}
    \textbf{Parameter}               &  \textbf{Prior range} \\
    \hline
    \hline
    $\omega_{\mathrm{b}}$   & [0.01875, 0.02625] \\
    \hline
    $\omega_{\mathrm{cdm}}$ & [0.05, 0.255]      \\
    \hline
    $h$                     & [0.64, 0.82]       \\
    \hline
    $n_s$                   & [0.84, 1.1]        \\
    \hline
    $S_8$                   & [0.1, 1.3]           \\
    \hline
    $c_{\mathrm{min}}$      & [2, 4]           \\
    \hline
    $A_{\mathrm{IA}}$       & [-6, 6]          \\
    \hline
    $D_{z_i}$               & $\mathcal{N}(\boldsymbol{\mu}, \boldsymbol{C})$          \\
    \hline
    \hline
    \end{tabular}
   \caption{Prior ranges for the KiDS-1000 cosmic shear band power analysis. Prior distributions are all taken to be uniform across these ranges, except for the $z$-shifts $D_{z_i}$, which are sampled from a correlated Gaussian prior with mean $\boldsymbol{\mu}$ and covariance $\boldsymbol{C}$ (see \citealt{Asgari20} for details on their values). Note that in this case we sample $S_8$ instead of $\mathrm{ln}10^{10}A_{\mathrm{s}}$. As explained in the main text, our emulators are trained on different choices of this sampling parameter, and one can always recover the parameter that is not directly sampled in a post-processing step, using Gaussian Process regression.}
  \label{tab:prior_kids1000}
\end{table}

\section{Results for CMB power spectrum emulator on larger range}
\label{app:cmb_large}
\begin{figure*}
        \centering
        \begin{subfigure}[b]{0.475\textwidth}
            \centering
            \includegraphics[width=\textwidth]{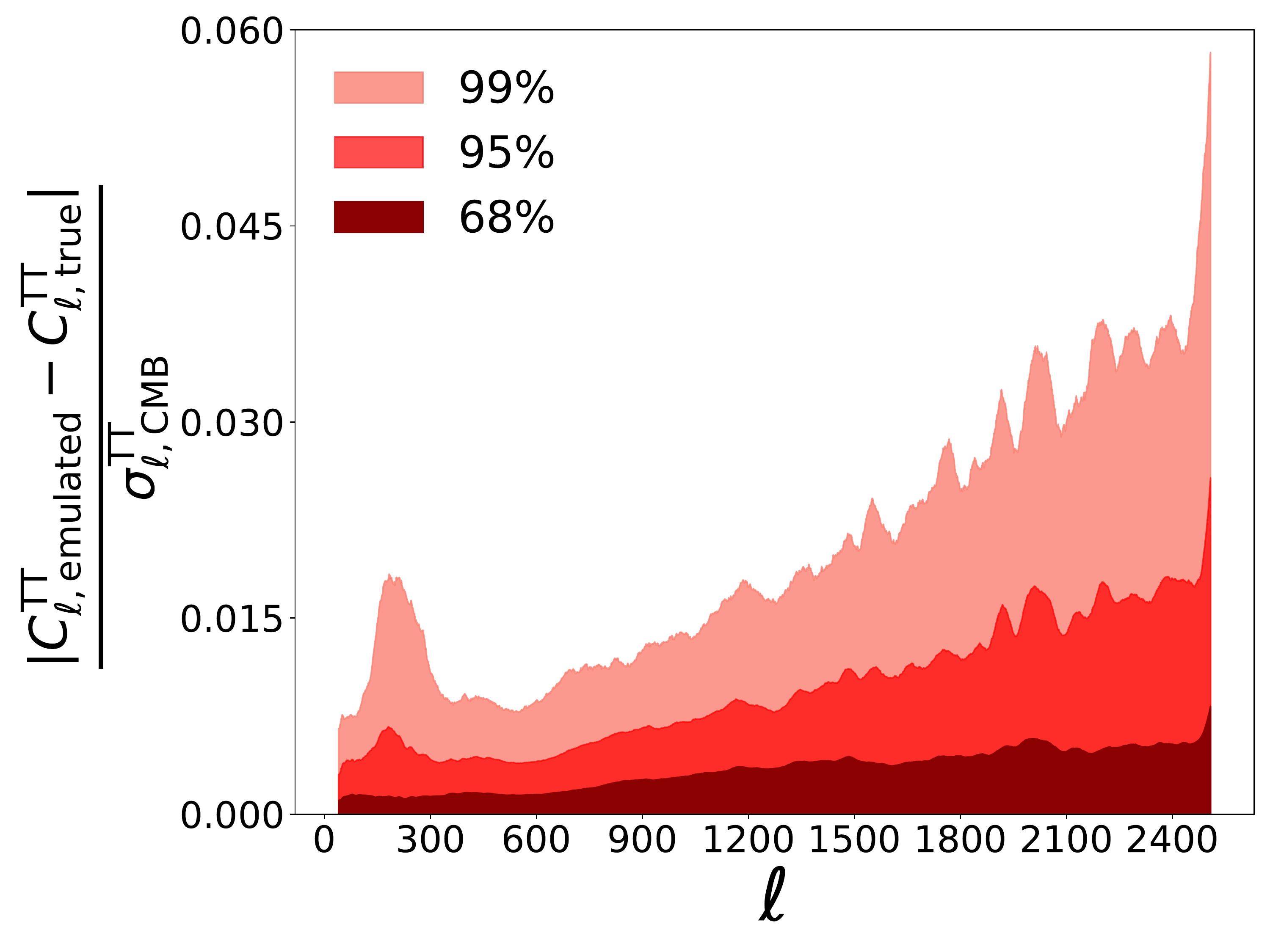}
            \caption[]%
            {{\small  $C_{\ell}^{\textrm{TT}}$}}    
            \label{fig:planck_tt_emu_wide}
        \end{subfigure}
        \hfill
        \begin{subfigure}[b]{0.475\textwidth}  
            \centering 
            \includegraphics[width=\textwidth]{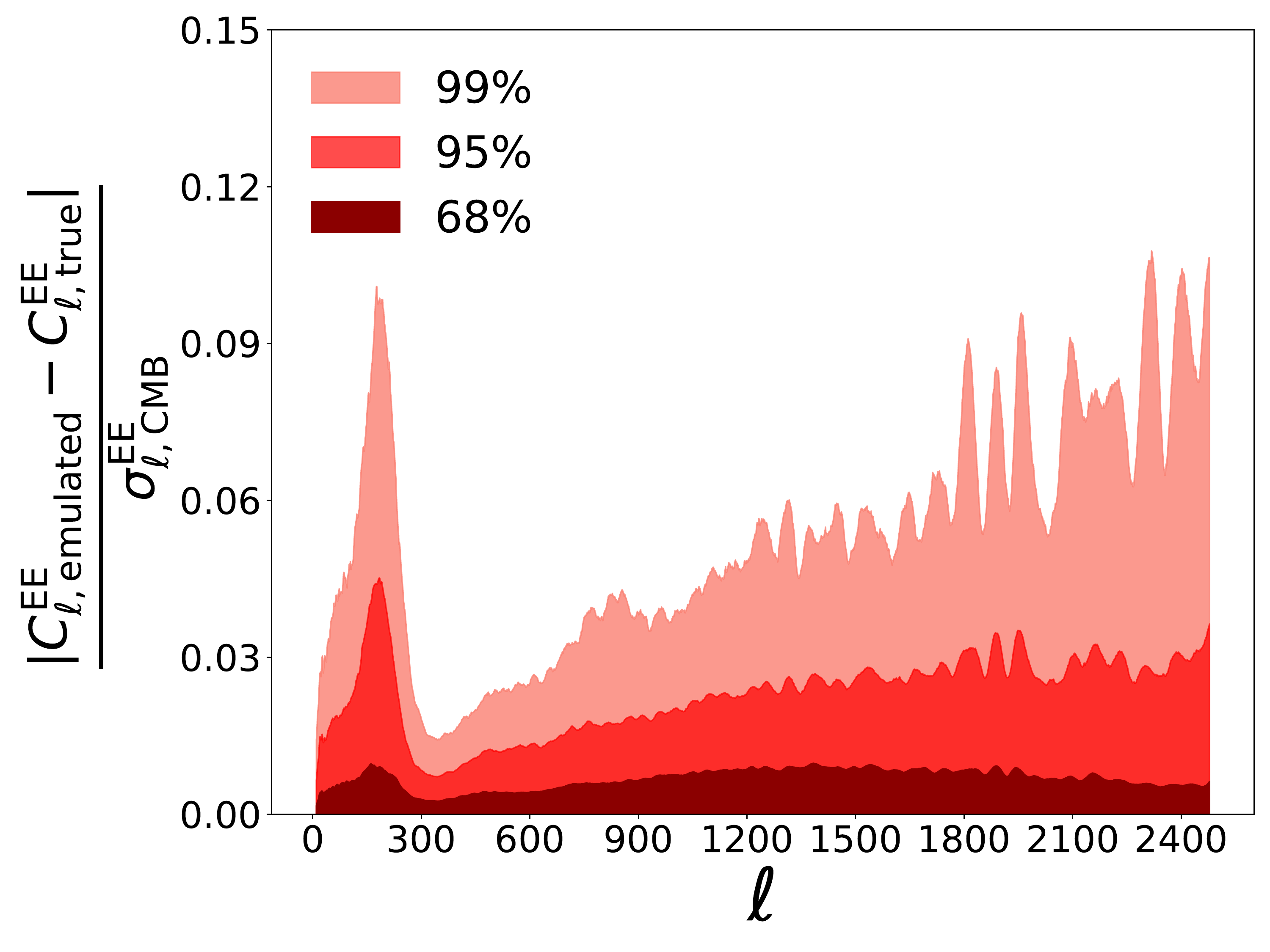}
            \caption[]%
            {{\small $C_{\ell}^{\textrm{EE}}$ }}    
            \label{fig:planck_ee_emu_wide}
        \end{subfigure}
        \vskip\baselineskip
        \begin{subfigure}[b]{0.475\textwidth}   
            \centering 
            \includegraphics[width=\textwidth]{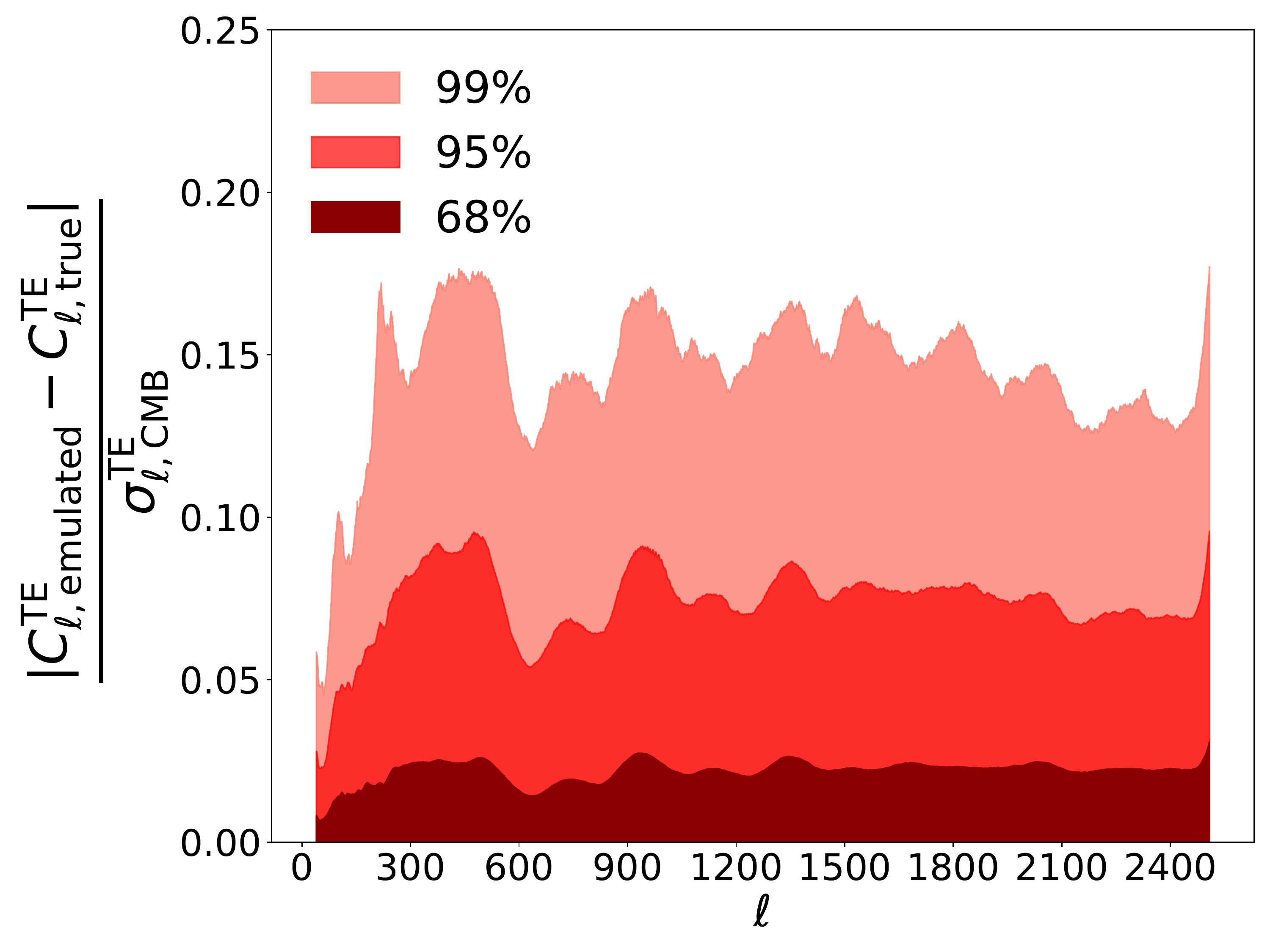}
            \caption[]%
            {{\small $C_{\ell}^{\textrm{TE}}$ }}    
            \label{fig:planck_te_emu_wide}
        \end{subfigure}
        \hfill
        \begin{subfigure}[b]{0.475\textwidth}   
            \centering 
            \includegraphics[width=\textwidth]{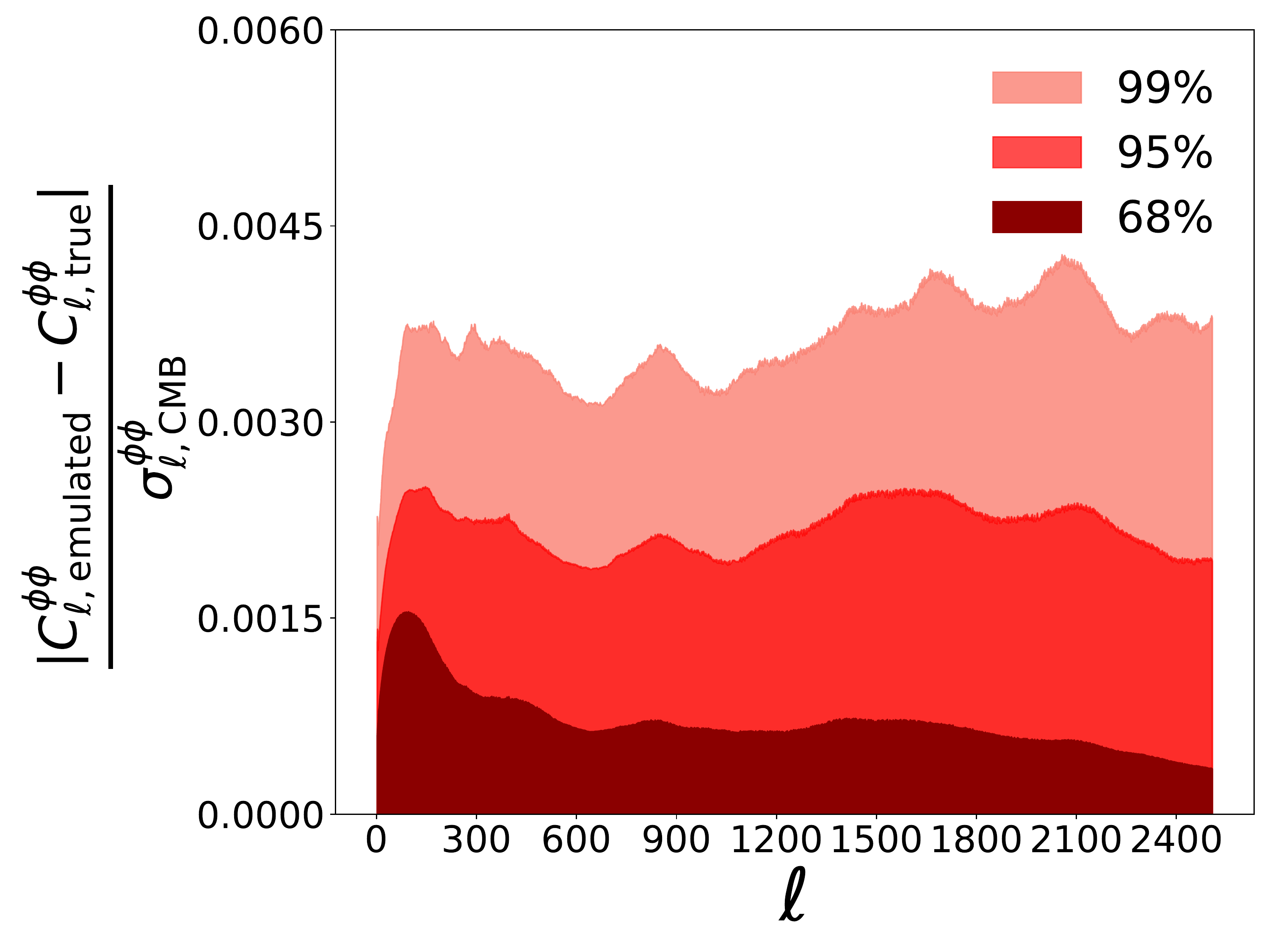}
            \caption[]%
            {{\small $C_{\ell}^{\phi \phi}$}}    
            \label{fig:planck_pp_emu_wide}
        \end{subfigure}
        \caption{CMB power spectra emulation accuracy on the test set sampled from the range indicated in Table~\ref{tab:emu_ranges} for a) the temperature power spectrum, b) the polarisation power spectrum, c) the temperature-polarisation cross power spectrum, d) the lensing potential power spectrum. The emulation error is defined with respect to both instrumental and statistical noise, and is defined in Eqs.~(\ref{eq:cl_cmb_error}-\ref{eq:cl_cmb_error_te}). \textit{Dark red}, \textit{red} and \textit{salmon} areas enclose the 68, 95 and 99 percentiles of the test set. Details of the neural models are reported in Appendix~\ref{app:nn}.}
        \label{fig:planck_emu_wide}
    \end{figure*}

In Fig.~\ref{fig:planck_emu_wide} we report the emulation accuracy of our CMB emulators applied to the test set sampled with cosmological parameters in the range of Table~\ref{tab:emu_ranges}. The performance reaches no more than 0.2$\sigma_{\textrm{CMB}}$ as defined in Eqs.~(\ref{eq:cl_cmb_error}-\ref{eq:cl_cmb_error_te}), which we verified corresponds to an adequate accuracy for Stage IV surveys, as we showed in the main body.

\bsp	
\label{lastpage}
\end{document}